\documentclass[a4paper,12pt]{article}
 \pdfoutput=1

\usepackage{graphicx}
\usepackage{hyperref}

\usepackage{a4wide}

\usepackage{amsmath,amssymb}
\usepackage{url}
\usepackage{cite}
\makeatletter

\def\section{\@startsection {section}{1}{\z@}{+3.0ex plus +1ex minus
  +.2ex}{2.3ex plus .2ex}{\normalsize\bf\boldmath}}
\def\subsection{\@startsection{subsection}{2}{\z@}{+2.5ex plus +1ex
minus +.2ex}{1.5ex plus .2ex}{\normalsize\bf\boldmath}}
\def\subsubsection{\@startsection{subsubsection}{3}{\z@}{+3.25ex plus
 +1ex minus +.2ex}{1.5ex plus .2ex}{\normalsize\it}}

\expandafter\ifx\csname mathrm\endcsname\relax\def\mathrm#1{{\rm #1}}\fi

\makeatother

\allowdisplaybreaks

\newcommand{\nus}{\nu_\mathrm{spin}^S}  
\newcommand{\nua}{\nu_\mathrm{ann}^{R,S}}  

\newcommand{\nuseps}{\nu_\mathrm{spin}^{S,\epsilon}}  

\allowdisplaybreaks
\numberwithin{equation}{section}

\begin{document}
\thispagestyle{empty}
\setcounter{page}{0}

\begin{flushright}
{\small
SI-HEP-2018-03 \\
TTK-18-02 \\
24 August 2018}
\end{flushright}

\vspace{\baselineskip}

\begin{center}
\textbf{\large Soft-gluon and Coulomb corrections to hadronic \\[0.02cm]
  top-quark pair production beyond NNLO}\\
\vspace{3\baselineskip}
{\sc  
Jan Piclum$^a$, Christian Schwinn$^b$
}\\
\vspace{0.7cm}
{\sl
${}^a$  
Theoretische Physik 1, Naturwissenschaftlich-Technische Fakult\"at,
Universit\"at Siegen,  Walter-Flex-Stra\ss e 3, 
57068 Siegen, Germany\\  \vspace{0.3cm}  
${}^b$Institut f\"ur Theoretische Teilchenphysik und   
Kosmologie,  
RWTH Aachen University, Sommerfeldstra\ss e 16,  D--52056 Aachen, Germany
}

\vspace*{1.2cm}
\textbf{Abstract}\\ 

\vspace{1\baselineskip}
\parbox{0.9\textwidth}{ We construct a resummation at partial
  next-to-next-to-next-to-leading logarithmic accuracy for hadronic
  top-quark pair production near partonic threshold, including
  simultaneously soft-gluon and Coulomb corrections, and use this
  result to obtain approximate next-to-next-to-next-to-leading order
  predictions for the total top-quark pair-production cross section at
  the LHC. We generalize a required one-loop potential in
  non-relativistic QCD to the colour-octet case and estimate the
  remaining unknown two-loop potentials and three-loop anomalous
  dimensions. We obtain a moderate correction of $1.5\%$ relative to
  the next-to-next-to-leading order prediction and observe a reduction
  of the perturbative uncertainty below $\pm 5\%$.
}
\end{center}
\newpage
\section{Introduction}

The total top-quark cross section at hadron colliders is a key
observable of the standard model that serves both as a probe of the
heaviest known elementary particle as well as a benchmark process of
perturbative Quantum Chromodynamics~(QCD). Comparison of measurements
at the Tevatron and LHC with precise theoretical predictions allows to
measure the top mass in a well-defined scheme, where recent
measurements have reached an accuracy of below
$2$~GeV~\cite{Aad:2014kva,Khachatryan:2016mqs}. Further applications
include determinations of the strong
coupling constant $\alpha_s$~\cite{Klijnsma:2017eqp,Chatrchyan:2013haa} and global
PDF fits, see e.g.~\cite{Ball:2014uwa,Harland-Lang:2014zoa,Alekhin:2017kpj}.
The experimental results from the LHC experiments have now reached an
impressive accuracy, with an uncertainty of better than $\pm 4\%$
(see~\cite{Hawkings:2017aiz} for a recent review), challenging the best
theoretical predictions based on next-to-next-to-leading order~(NNLO)
QCD~\cite{Baernreuther:2012ws,Czakon:2012zr,Czakon:2012pz,Czakon:2013goa}
supplemented with next-to-next-to-leading logarithmic~(NNLL) soft-gluon
resummation~\cite{Beneke:2009rj,Czakon:2009zw,Beneke:2011mq,Cacciari:2011hy,
  Czakon:2011xx, Beneke:2012wb}. 

Since a full computation of top-quark pair production at N$^3$LO accuracy in
QCD is currently out of reach, attempts to reduce the perturbative
uncertainties further rely on resummation methods. In this article
we explore prospects for a combined resummation of soft-gluon and Coulomb
gluon effects at N$^3$LL accuracy in the framework
of~\cite{Beneke:2009rj,Beneke:2010da} and provide expressions for the total
cross section at N$^3$LO in the partonic threshold limit $\beta=\sqrt{1-4
  m_t^2/\hat s}\to 0$.  Coulomb corrections arise from the exchange of gluons
between the slowly moving top quarks and lead to corrections of the form
$(\alpha_s/\beta)^k$, which can be resummed to all orders using Green-function
methods in non-relativistic QCD~(NRQCD).  Counting both the logarithmic
soft-gluon corrections $\alpha_s \ln\beta$ and the Coulomb corrections
$\alpha_s/\beta$ arising at each order in perturbation theory as quantities of
order one, a combined resummation of these effects rearranges the perturbative
series of the partonic cross section into the schematic
form~\cite{Beneke:2009rj}
\begin{align}
\label{eq:syst}
\hat{\sigma} &\propto \hat\sigma^{(0)} 
\sum_{k=0} \left(\frac{\alpha_s}{\beta}\right)^k \!\!\!
\exp\Big[\underbrace{\ln\beta\,g_0(\alpha_s\ln\beta)}_{\mbox{(LL)}}+ 
\underbrace{g_1(\alpha_s\ln\beta)}_{\mbox{(NLL)}}+
\underbrace{\alpha_s g_2(\alpha_s\ln\beta)}_{\mbox{(NNLL)}}+
\underbrace{\alpha_s^2 g_3(\alpha_s\ln\beta)}_{\mbox{(N$^3$LL)}}+\ldots\Big]
\nonumber\\[0.2cm]
& \quad\times
\left\{1\,\mbox{(LL,NLL)}; \alpha_s,\beta \,\mbox{(NLL',NNLL)}; 
\alpha_s^2,\alpha_s\beta,\beta^2 \,\mbox{(NNLL',N$^3$LL)};
\ldots\right\}.
\end{align}
We also indicated a modified counting N$^n$LL', where the fixed-order
corrections in the second line are included at one order higher than in the
``unprimed'' counting.  The combined resummation of soft and Coulomb corrections
at NNLL accuracy~\cite{Beneke:2011mq} was implemented in the program
\texttt{topixs}~\cite{Beneke:2012wb}, whose current version includes the
matching to the complete NNLO
corrections~\cite{Baernreuther:2012ws,Czakon:2012zr,Czakon:2012pz,Czakon:2013goa}. NNLO+NNLL
soft-gluon resummation with the Mellin-transform method~\cite{Cacciari:2011hy}
with a fixed-order treatment of Coulomb corrections was implemented in
the program~\texttt{top++}~\cite{Czakon:2011xx}, which further includes the
$\mathcal{O}(\alpha_s^2)$ constant terms in the resummation that are part of
the NNLL' corrections in~\eqref{eq:syst}.  Other NNLL resummations based on
pair-invariant mass or single-particle-inclusive
observables~\cite{Ahrens:2010zv,Ahrens:2011mw,Broggio:2014yca} also do not
include the Coulomb corrections.  In contrast, a combination of resummed
Coulomb corrections with fixed-order soft corrections was performed
in~\cite{Hagiwara:2008df,Kiyo:2008bv,Sumino:2010bv}.  

Top quarks are not dominantly produced at threshold at the LHC, so
threshold-enhanced corrections do not necessarily require resummation and
fixed-order perturbation theory is expected to be adequate for the total cross
section.\footnote{This holds up to N$^4$LO where the presence of a
  $\alpha_s^4/\beta^4$ correction renders the convolution with the parton
  luminosity unintegrable, so resummation of Coulomb corrections may be
  required despite a small numerical effect.}  To the extent that these terms
nevertheless constitute a significant part of the full higher-order
corrections, it is therefore justified to expand a resummed prediction to a
fixed order. In this way, N$^n$LL resummation predicts all N$^n$LO terms that
become singular for $\beta\to 0$, while a constant correction relative to the
Born cross section $\hat\sigma^{(0)}$ is further included at N$^n$LL'.
Experience at
NNLO~\cite{Moch:2008qy,Beneke:2009ye,Kidonakis:2010dk,Ahrens:2011px} indicates
that the singular terms provide a feasible approximation to the full result if
corrections due to possibly sizeable non-singular terms are estimated in a
sufficiently conservative way.  For instance, the prediction
$\sigma^{\mathrm{NNLO}_{\mathrm{app}}}_{t\bar t}(7\,\mathrm{ TeV})=
161.1{}^{+11.4}_{-10.9}{}^{+4.7}_{-4.7}$~pb was obtained
in~\cite{Beneke:2011mq}  using the partonic cross sections calculated in~\cite{Beneke:2009ye}, where the first error refers to the scale uncertainty
and the second one estimates terms beyond the threshold approximation.  This
result is consistent with the full NNLO calculation
$\sigma^{\mathrm{NNLO}}_{t\bar t}(7\,\mathrm{ TeV})=
167.0^{+6.7}_{-10.7}$~pb~\cite{Czakon:2013goa} and improves the accuracy
compared to the NLO prediction $\sigma^{\mathrm{NLO}}_{t\bar t}(7\,\mathrm{
  TeV})= 158.0^{+19.5}_{-21.2}$~pb. This motivates the construction of
approximate N$^3$LO corrections, which should be based on N$^3$LL accuracy to
obtain all threshold-enhanced terms.

At present, a complete N$^3$LL resummation for top-quark production is not
feasible, since some three- and four-loop coefficients in the resummation
function $g_3$ are not known for the colour-octet case.  In addition, starting
from NNLL logarithmic corrections arise also from Coulomb corrections, which
are governed by renormalization-group equations in
NRQCD~\cite{Pineda:2001et,Hoang:2002yy}, with anomalous dimensions only
fully known for colour-singlet states.\footnote{In the NNLL calculation
  of~\cite{Beneke:2011mq,Beneke:2012wb} these terms are included at fixed
  order.}  Furthermore, the power corrections $\sim \beta^2$ in the curly
brackets in~\eqref{eq:syst} must be controlled in order to achieve N$^3$LL
accuracy according to the counting $\alpha_s\sim \beta$.
This makes a complete resummation at this accuracy conceptually and
technically challenging.

In this paper, we construct a partial N$^3$LL approximation by including all
currently known ingredients of N$^3$LL soft-gluon
resummation~\cite{Becher:2007ty,Czakon:2013hxa,Baernreuther:2013caa} and the
Coulomb corrections from a recent calculation of $e^-e^+\to t\bar t$ at
N$^3$LO in potential
NRQCD~\cite{Beneke:2013jia,Beneke:2015kwa,Beneke:2016kkb}.  We further include
several sources of power-suppressed corrections, i.e.\ contributions of P-wave
production channels and the combination of Coulomb corrections and a so-called next-to-eikonal logarithm.  In
this way we obtain all threshold enhanced N$^3$LO terms for the colour-singlet
state and are in a position to estimate the uncertainty due to missing
ingredients for colour-octet states. We also include an $\mathcal{O}(\alpha_s^3)$ contribution to the
Coulomb corrections that does not follow from the
straightforward expansion of the well-known Sommerfeld factor for stable
particles~\cite{Beneke:2016jpx}.

We compare our results to other recent works on N$^3$LO
effects in top-quark pair production.  In~\cite{Beneke:2011mq} the expansion
of the NNLL cross section was used, which is not sufficient to predict
terms of the form $\alpha_s^3\{\ln^{2,1}\beta,1/\beta \} $ exactly.  In the
momentum-space resummation approach~\cite{Becher:2006nr,Becher:2006mr,Becher:2007ty} used
in~\cite{Beneke:2011mq}, this incomplete knowledge manifests itself in a
residual dependence on unphysical hard and soft scales which results in a
sizeable uncertainty.  A prediction based on
NNLL resummation in one-particle inclusive kinematics was made in~\cite{Kidonakis:2014isa}. No attempt is made to
estimate the systematic uncertainties of this
approximation. In~\cite{Muselli:2015kba} threshold resummation for the
invariant-mass distribution is combined with information about the large-$x$
limit. However, only the gluon-initiated subprocess is considered.

This paper is organized as follows: In Section~\ref{sec:frame} we outline the
framework of our calculation and discuss consequences of collinear
factorization and renormalization-group invariance for N$^3$LO corrections.
In Section~\ref{sec:n3ll} we discuss in detail the inputs of the partial
N$^3$LL resummation while the approximate N$^3$LO results for partonic cross
sections and predictions for the LHC are presented in
Section~\ref{sec:results}. Explicit results for the renormalization group
evolution of the two-loop soft and hard functions are given in
Appendix~\ref{app:input}. The generalization of some contributions to the
potential corrections to the required spin and colour states are discussed in
Appendix~\ref{app:potential}. Explicit results for the factorization-scale
dependent contributions to the partonic cross sections are collected in Appendix~\ref{app:scaling}.

\section{Factorization and resummation framework}
\label{sec:frame}

In this section we collect results for the factorization and renormalization
scale dependence of N$^3$LO corrections, discuss the colour and
spin states of top-quark production near threshold, and outline the
resummation formalism used for the combined soft and Coulomb gluon resummation
at NNLL~\cite{Beneke:2009rj,Beneke:2010da,Beneke:2011mq,Beneke:2012wb}.
\subsection{Setup of the perturbative calculation}
\label{sec:setup}

The total hadronic cross section for the production of a $t\bar t+X$ 
final state in collisions of hadrons $N_{1,2}$ with centre-of-mass energy 
$s$ is obtained from the convolution of the partonic cross section with the
parton luminosity,
\begin{equation}
\sigma_{N_1 N_2\to t\bar t X}(s)=
\sum_{p,p'=q,\bar q,g}\,\int_{4 m_t^2/s}^1 \!d\tau\,L_{pp'}(\tau,\mu_f)
\,\hat\sigma_{pp'} (s \tau,\mu_f,\mu_r)\,,
\label{eq:sig-had}
\end{equation}
where the latter is defined in terms of the parton distributions 
functions~(PDFs)
\begin{equation}
\label{eq:lumi}
L_{p p^\prime}(\tau,\mu) = \int_0^1 dx_1
dx_2\,\delta(x_1 x_2 - \tau) \,f_{p/N_1}(x_1,\mu)f_{p^\prime/N_2}(x_2,\mu)\,.
\end{equation}
The perturbative expansion of the partonic cross section in the strong
coupling constant is conveniently expressed in terms of corrections relative to the Born
cross section,
\begin{equation}
\hat \sigma_{pp'}(\hat s,m_t,\mu_f,\mu_r) =
\hat \sigma^{(0)}_{pp'}(\hat s,m_t,\mu_r) \Bigg\{ 1
+  \sum_{n=1}^\infty\left(\frac{\alpha_s(\mu_r)}{4\pi}\right)^n
\hat \sigma^{(n)}_{pp'}(\hat s,m_t\,\mu_f,\mu_r) \biggr\}.
\label{eq:sigma-series}
\end{equation}
Up to $\mathcal{O}(\alpha_s^2)$, the corrections $\hat
\sigma^{(n)}_{pp'}$ are known exactly for all partonic
channels~\cite{Baernreuther:2012ws,Czakon:2012zr,Czakon:2012pz,Czakon:2013goa}. The
numerical predictions have been parameterized by fitting functions and implemented in the most recent versions of the programs
\texttt{top++}~\cite{Czakon:2011xx}, \texttt{HATHOR}~\cite{Aliev:2010zk}, and \texttt{topixs}~\cite{Beneke:2012wb}.
The goal of this paper is to predict the partonic
cross sections in the leading production channels $pp'=gg,q\bar q$ up to
$\mathcal{O}(\alpha_s^3)$ in the threshold limit $\hat s\to 4 m_t^2$. 

 The 
dependence of the cross section~\eqref{eq:sigma-series} on the renormalization scale $\mu_r$  can be reconstructed from a calculation of the
corrections $ \hat \sigma^{(n)}_{pp'}(\hat s,m_t\,\mu_f)$ for
$\mu_f=\mu_r$ by re-expressing the result as an expansion in
$\alpha_s(\mu_r)$. In this way, the renormalization-scale dependence of the
N$^{3}$LO cross section is obtained as
\begin{align}
\hat \sigma^{(3)}_{pp'}(\hat s,m_t\,\mu_f,\mu_r) &=
 \hat \sigma^{(3)}_{pp'}(\hat s,m_t\,\mu_f)
  -8\beta_0 \ln\left(\frac{\mu_f}{\mu_r}\right)\,
 \hat \sigma^{(2)}_{pp'}(\hat s,m_t\,\mu_f)\nonumber\\
&\qquad +6\left[4\beta_0^2\ln^2\left(\frac{\mu_f}{\mu_r}\right)-
     \beta_1 \ln\left(\frac{\mu_f}{\mu_r}\right) \right] 
 \hat \sigma^{(1)}_{pp'}(\hat s,m_t\,\mu_f)\nonumber\\
  &\qquad -4\left[8\beta_0^3\ln^3\left(\frac{\mu_f}{\mu_r}\right)-
     7\beta_0\beta_1 \ln^2\left(\frac{\mu_f}{\mu_r}\right)
     +\beta_2\ln\left(\frac{\mu_f}{\mu_r}\right) \right] .
\label{eq:mur}
\end{align}
In a strict threshold expansion, the limit $\hat s\to 4m_t^2$ of the partonic
cross sections on the right-hand side is used and 
the terms in the last line are dropped, since they contribute to the terms $\mathcal{O}(\beta^0)$
relative to the Born cross section.

The factorization-scale dependence of the cross section
can be obtained from results at lower perturbative order by exploiting
the known factorization-scale dependence of the PDFs, which implies
the evolution equation of the partonic cross section
\begin{equation}
\label{eq:dmu-sigma}
  \frac{d}{d\ln\mu}\hat \sigma_{pp'}(\hat s,m_t,\mu)
  =-\sum_{\tilde p,\tilde p'}\int_{4m_t^2/\hat s}^1\frac{dx}{x}\left(P_{p/\tilde
      p}(x)+ P_{p'/\tilde p'}(x)\right)
  \hat \sigma_{\tilde p \tilde p'}(x\hat s,\mu),
\end{equation}
where  $P_{p/\tilde p\,}(x)$ are the Altarelli-Parisi 
splitting functions for the
splitting of a parton $p$ into a parton $\tilde p$.  
In the threshold limit  it is consistent to use the $x\to 1$
limit of the splitting functions for a parton $p$ in the colour representation
$r$,\footnote{
Here some care has to be taken since subleading terms in the $x\to 1$ limit
can be enhanced by the Coulomb corrections.  In can be checked, however, that
the leading correction to the N$^3$LO cross section from the $\mathcal{O}(1-x)$
term in~\eqref{eq:splitting} is of order $\alpha_s^3\beta^2$, and therefore
beyond N$^3$LL.
} which is given in terms of the light-like cusp-anomalous dimension
$\Gamma^r_{\text{\text{cusp}}}$ and a subleading anomalous dimension $\gamma^{\phi,r}$
\begin{equation}
\label{eq:splitting}
P_{p/\tilde p\,}(x)
=\left(2 \Gamma^r_{\text{\text{cusp}}}(\alpha_s)
\frac{1}{[1-x]_+}+2 \gamma^{\phi,r}(\alpha_s) \delta(1-x)\right)
\delta_{p\tilde p}+\mathcal{O}(1-x) \, .
\end{equation}
The anomalous dimensions are all known at least up to three-loop
level~\cite{Moch:2004pa} and summarized in the conventions used here in~\cite{Becher:2007ty,Ahrens:2008nc}.

In the higher-order corrections to the cross section, the dependence on the
factorization scale can be made explicit by the decomposition
\begin{equation}
\hat \sigma^{(n)}_{pp'}(\hat s,m_t\,\mu_f)=
 \sum_{m=0}^n f^{(n,m)}_{pp'}(\rho) \ln^m\left(\frac{\mu_f}{m_t}\right),
\label{eq:scaling}
\end{equation}
with the so-called scaling functions $f^{(n,m)}_{pp'}$ and the variable $\rho=\frac{4m_t^2}{\hat s}$. 
For S-wave production channels with $\hat
\sigma^{(0)}\propto \alpha_s^2(\mu_f)\beta$ it is useful to define modified
scaling functions\footnote{These definitions are related to~\cite{Beneke:2009ye} by $g^{(n,m)}_{pp'}=2^{n+m}s^{(n,m)}_{pp'}$ while
  our convention for the splitting function $P^{(n)}$ is $2^{n+2}$ times the
  one used there.} 
\begin{equation}
   g_{pp'}^{(n,m)}(\rho)=\beta f_{pp'}^{(n,m)}(\rho).
\end{equation}
 In
the threshold limit $\rho\to 1$ we obtain the 
results for the N$^3$LO scaling functions
\begin{align}
  g_{pp}^{(3,3)}&=\frac{1}{3}
\left[8\beta_{0}\,g_{pp}^{(2,2)}-
2g^{(2,2)}_{pp}\otimes P^{(0)}_{p/p} \right],\\
 g_{pp}^{(3,2)}&=
4\beta_{0}\,g_{pp}^{(2,1)}+3\beta_{1}\,g_{pp}^{(1,1)}-
g^{(2,1)}_{pp}\otimes P^{(0)}_{p/p}
-g^{(1,1)}_{pp}\otimes P^{(1)}_{p/p}, \\
 g_{pp}^{(3,1)}&=
8\beta_{0}\,g_{pp}^{(2,0)}+6\beta_{1}\,g_{pp}^{(1,0)}+4\beta_{2}\,g_{pp}^{(0,0)}
\nonumber\\
&\quad -g^{(2,0)}_{pp}\otimes P^{(0)}_{p/p}
-g^{(1,0)}_{pp}\otimes P^{(1)}_{p/p} 
-g^{(0,0)}_{pp}\otimes P^{(2)}_{p/p} ,
\end{align}
where the threshold limit of the lower-order scaling functions is consistently
used on the
right-hand side.
The convolutions are
defined as
\begin{equation}
\label{eq:def-conv}
(g\otimes P)(\rho)=\int_\rho^1 dx_1 dx_2  g(x_1) P(x_2)\delta(x_1x_2-\rho)
\end{equation}
and the conventions for the coefficients $\beta_n$ and $P_{p/p}^{(n)}$ of
the perturbative expansion of the beta function and the splitting functions are
spelled out in~\eqref{eq:beta-alpha} and~\eqref{eq:p-alpha}.

\subsection{Top-quark production channels near partonic threshold}

The resummation of soft-gluon and Coulomb corrections requires to decompose
the partonic production cross sections into contributions with definite colour
and spin states of the top-quark pair, $\sigma^{(0),R,{}^{2S+1}L_J}_{pp'}$. For the colour representations, the familiar decomposition of the
tensor product $3\otimes\bar 3=1+8$ of the $SU(3)$ representations is used,
where for gluon initial states the symmetric and antisymmetric colour channels
with respect to gluon exchange, $8_s$ and $8_a$, are distinguished. For the
spin states, the spectroscopic notation for orbital angular momentum
$L=\mathrm{S},\mathrm{P},\dots$, spin $S$ and total angular momentum $J$ is
used.

In the quark-antiquark induced production channel, the top-antitop pair is
dominantly produced in a colour-octet ${}^3\mathrm{S}_1$ state, while the
dominant production channel in gluon fusion is given by a colour-singlet or
symmetric octet ${}^1\mathrm{S}_0$ state.  The threshold limit of the leading
order S-wave cross sections is given by
\begin{subequations}
\label{eq:born}
\begin{align}
\sigma^{(0),8,{}^3\mathrm{S}_1}_{q\bar q} &= \pi{(N_c^2-1)\over 2N_c^2}
\frac{\alpha_s^2(\mu^2)}{\hat s}\beta ,\label{eq:qqbar-exp}  \\
\sigma^{(0),8_s,{}^1\mathrm{S}_0}_{gg} &= \pi{(N_c^2-4)\over 2N_c(N_c^2-1)}
\frac{\alpha_s^2(\mu^2)}{ \hat s} \beta, \\
\sigma^{(0),1,{}^1\mathrm{S}_0}_{gg} &= \pi{1\over N_c(N_c^2-1)}
\frac{\alpha_s^2(\mu^2)}{\hat s}\beta,\label{eq:gg-S0} 
\end{align}
\end{subequations}
where we have left an overall factor $4m_t^2/\hat s$ unexpanded.
The spin label in these expressions will be dropped if no confusion can arise.
For the counting $\alpha_s\sim \beta$ used in~\eqref{eq:syst} also subleading
terms of $\mathcal{O}(\beta^2)$ in the threshold expansion of the Born cross
section must be taken into account at N$^3$LL accuracy.  For the S-wave
production processes~\eqref{eq:born}, these terms are given by a correction
factor 
$1-\frac{1}{3}\beta^2$. 
At the same order, there are contributions to the
total cross section from
P-wave production channels ${}^3\mathrm{P}_0$, and
${}^3\mathrm{P}_2$ (see
e.g.~\cite{Petrelli:1997ge}) with the
contributions to the total production cross section
\begin{subequations}
\label{eq:pwave}
 \begin{align}
    \sigma^{(0),R,{}^3\mathrm{P}_0}_{gg}&=
    \sigma^{(0),R,{}^1\mathrm{S}_0}_{gg}\; \beta^2, \\
      \sigma^{(0),R,{}^3\mathrm{P}_2}_{gg}&=
    \sigma^{(0),R,{}^1\mathrm{S}_0}_{gg} \; \frac{4}{3}\beta^2 .
  \end{align}
\end{subequations}
 Since the second and third Coulomb corrections
differ for S-wave and P-wave production channels (see
e.g.~\cite{Falgari:2012hx}) it is necessary to distinguish the different
angular momentum states contributing to the subleading Born
contributions.

In addition to the colour-singlet and symmetric octet channels, there
is also a kinematically suppressed contribution from the 
 antisymmetric colour-octet  channel  
\begin{equation}
  \sigma^{(0),8_a,{}^1\mathrm{S}_0}_{gg} = \pi \frac{N_c}{6(N_c^2-1)}
\frac{\alpha_s^2(\mu^2)}{\hat s} \beta^3.
\label{eq:born-8A}
\end{equation}
As mentioned in~\cite{Baernreuther:2013caa} and discussed in the context of
the violation of the Landau-Yang theorem in
QCD~\cite{Beenakker:2015mra,Cacciari:2015ela}, this suppression is not the
signal of P-wave production but the result of an accidental cancellation in
the Born S-wave matrix element. 

\subsection{Combined soft and Coulomb resummation at NNLL}
\label{sec:resum}
Threshold resummation of soft
logarithms~\cite{Sterman:1986aj,Catani:1989ne} was first established for top-quark pair
production at
NLL~\cite{Kidonakis:1997gm,Bonciani:1998vc}
and more recently at NNLL for the total cross section ~\cite{Beneke:2009rj,Czakon:2009zw,Beneke:2011mq,Cacciari:2011hy} 
and  pair-invariant mass or single-particle-inclusive
observables~\cite{Ahrens:2010zv,Ahrens:2011mw,Broggio:2014yca}. 
The combination of soft-gluon and Coulomb-gluon resummation up to
NNLL in the combined counting~\eqref{eq:syst} was shown
in~\cite{Beneke:2009rj,Beneke:2010da} for pairs of heavy coloured particles
produced in an S-wave state. This method was applied to top-quark production
in~\cite{Beneke:2011mq,Beneke:2012wb}. The basis for the joint soft and
Coulomb resummation is the factorization of the total partonic production
cross section into a potential function $J$, a hard function $H$, and a soft
function $W$~\cite{Beneke:2010da}:
\begin{equation}
\label{eq:fact}
  \hat\sigma_{pp'}(\hat s,\mu_f)
=\frac{4m_t^2}{\hat s} \sum_{R=1, 8}\sum_{S=0}^1H^{R,S}_{pp',i}(m_t,\mu_f)
\;\int d \omega\;
J_{R}^{S}(E-\frac{\omega}{2})\,
W^{R}_i(\omega,\mu_f)\, ,
\end{equation}
which was derived using the leading-power Lagrangians of soft-collinear
effective theory~(SCET) and potential non-relativistic QCD~(PNRQCD).  Here
$E=\sqrt{\hat s}-2 m_t$ is the energy relative to the production
threshold. The sum is over the colour representations $R$ of the final state
top-pair system, i.e.  the colour-singlet and octet states, and the total spin
$S=0,1$. The index $i$ denotes the colour basis for the hard scattering
$pp'\to t\bar t$~\cite{Beneke:2009rj}.
The leading-order hard function for the S-wave scattering process $pp'\to
(t\bar t)_R^{{}^{2S+1}\mathrm{S}_J}$ for the colour state $R$ and spin state $S$ is
related to the partonic Born cross sections in the corresponding channel at
threshold~\eqref{eq:born} according to
\begin{equation}
\label{eq:sigma-hard}
  \hat\sigma_{pp'}^{(0),R,{}^{2S+1}\mathrm{S}_J}(\hat s,\mu_f)
\underset{\hat s\to 4m_t^2}{\approx}
\frac{\beta m_t^2}{2\pi} H^{R,S(0)}_{pp'}(\mu_f) \, . 
\end{equation}
In~\eqref{eq:fact} we have made the conventional prefactor $4m_t^2/\hat s$
explicit that is implicitly included in the hard function in~\cite{Beneke:2010da}.
 The soft function $W_i^{R}$
in~\eqref{eq:fact} is defined by a time-ordered expectation value of Wilson
lines and corresponds to the $2\to 1$ process of the production of a heavy
particle in the colour representation $R$~\cite{Beneke:2009rj}.  For resummation of threshold
logarithms the momentum-space method~\cite{Becher:2006nr,Becher:2006mr,Becher:2007ty} is used, where the soft and hard
functions are evolved from a soft
scale $\mu_s\sim m_t\beta^2$ and a hard scale $\mu_h\sim 2m_t$ to the
factorization scale $\mu_f\sim m_t$ using
renormalization-group equations~(RGEs) summarized in Section~\ref{sec:n3ll}
below.  The potential function is given in terms of the imaginary part of the
Coulomb Green function in non-relativistic QCD, which resums ladder diagrams
with Coulomb-gluon exchange to all orders in $\alpha_s$.  The resulting
resummed cross section is of the form~\cite{Beneke:2010da}
\begin{equation}  
  \label{eq:resum-sigma}  
\begin{aligned}  
\hat\sigma^{\text{res}}_{pp'}(\hat s,\mu_f)=&  
\frac{4m_t^2}{\hat s}\sum_{R,S}  
H_{pp',i}^{R,S}(m_t,\mu_h)\,  
U_{i}^R(\mu_h,\mu_s,\mu_f)  
\left(\frac{2m_t}{\mu_s}\right)^{-2\eta} \\  
&\times  
\tilde{s}_i^{R}(\partial_\eta,\mu_s)  
\frac{e^{-2 \gamma_E \eta}}{\Gamma(2 \eta)}\,  
\int_0^\infty \!\!\!\!\! d \omega   
\;\frac{ J^S_{R_{\alpha}}(E-\tfrac{\omega}{2})}{\omega}   
\left(\frac{\omega}{\mu_s}\right)^{2 \eta}.  
\end{aligned}  
\end{equation}  
Here the Laplace transformation  of the 
soft function,  
\begin{equation}
\label{eq:soft-laplace}  
 \tilde  s_i^{R}(\rho,\mu)=\int_{0}^{\infty} d \omega e^{-s \omega}  
\,  W_i^{R}(\omega,\mu) ,
\end{equation}
was introduced, where $s = 1/(e^{\gamma_E} \mu e^{\rho/2})$.
 After carrying out the differentiations with respect to $\eta$ in~\eqref{eq:resum-sigma}, this variable is identified with
a resummation function which contains single logarithms,
$\alpha_s\ln(\mu_s/\mu_f)$, while the
resummation function $U_i$ sums the Sudakov double logarithms
$\alpha_s\ln^2\frac{\mu_h}{\mu_f}$ and $\alpha_s\ln^2\frac{\mu_s}{\mu_f}$. The
precise definitions of these functions for the case of heavy-particle pair
production are given in~\cite{Beneke:2010da} and the expansions required for
N$^3$LL accuracy can be found in~\cite{Becher:2007ty}.
For the NNLL resummation carried out in~\cite{Beneke:2011mq}, the NLO
hard~\cite{Hagiwara:2008df,Czakon:2008cx} and soft~\cite{Beneke:2009rj}
functions have been used,
together with the three-loop cusp-anomalous dimension~\cite{Moch:2004pa} and
the remaining anomalous dimensions in the evolution equations at two-loop
accuracy.  In the Coulomb sector, using the NLO potential function quoted
in~\cite{Beneke:2011mq} resums all corrections of the form
$(\alpha_s/\beta)^k$ and $\alpha_s\times (\alpha_s/\beta)^k$. This was
supplemented by a leading resummation of logarithms by using a running
Coulomb scale, and the inclusion of the leading so-called non-Coulomb
correction~\cite{Beneke:2009ye,Beneke:2011mq,Beneke:2016kvz}, which give rise to a tower of terms of the form
$\alpha_s^2\ln\beta\times (\alpha_s/\beta)^k$.

\section{Towards  N$^3$LL resummation}
\label{sec:n3ll}
In this section we collect the input required for N$^{3}$LL resummation
according to the systematics defined in~\eqref{eq:syst} and identify missing
ingredients, whose effect will be estimated in the phenomenological results. 

For soft-gluon resummation alone, the required ingredients
to increase the logarithmic accuracy are well-defined, but not fully
available for N$^3$LL resummation.
For N$^n$LL accuracy according
to~\eqref{eq:syst}, the cusp-anomalous dimension has to be known at $(n+1)$-loop order, the
remaining anomalous dimensions in the evolution equations of the soft and
hard functions to $n$-loop order. The fixed-order soft and hard functions have
to be known to N$^{n-1}$LO (N$^n$LO) accuracy for the N$^n$LL (N$^n$LL')
approximation.  For top-quark production, NNLO soft and hard functions can be
obtained from results
of~\cite{Belitsky:1998tc,Becher:2007ty,Czakon:2013hxa,Baernreuther:2013caa} as
discussed in Sections~\ref{sec:hard} and~\ref{sec:soft}.  The four-loop cusp
anomalous dimension was recently computed for the quark
case~\cite{Moch:2017uml} but is still unknown for gluons. However, this
affects only the full N$^3$LL resummation but not the expansion to N$^3$LO
considered in this paper.  Further, the three-loop soft anomalous dimension
is known for the massless~\cite{Almelid:2015jia,Almelid:2017qju} but
not the massive case. 

Concerning the corrections in the Coulomb sector, the potential function for
the colour-singlet, spin-triplet case is known at the required accuracy from a
calculation of $e^-e^+\to t\bar
t$~\cite{Beneke:2013jia,Beneke:2015kwa,Beneke:2016kkb}.  This requires the
insertion of $\mathcal{O}(\alpha_s^2,\alpha_s\beta,\beta^2)$ suppressed
potentials, which are not fully known for colour-octet states. In
Section~\ref{sec:nnlo-pot} we obtain the expansion of the potential function
to $\mathcal{O}(\alpha_s^3)$. We partially generalize the result to general
spin and the colour-octet case and estimate the remaining unknown colour-octet
contributions.

However, the presence of Coulomb corrections near the total production threshold $\hat
s\to 4 m_t^2$ complicates the extension to higher accuracy compared to
resummations for other kinematical threshold definitions such as the
pair-invariant mass or single-particle-inclusive
observables~\cite{Ahrens:2010zv,Ahrens:2011mw,Broggio:2014yca}.
 The enhancement of Coulomb corrections by negative
powers of $\beta$ requires to control the corrections involving positive
powers of $\beta$ in the curly brackets in~\eqref{eq:syst} to achieve the
desired accuracy using the counting $\alpha_s\sim \beta$.
In the
effective-theory framework of~\cite{Beneke:2009rj,Beneke:2010da} used to
derive the factorization~\eqref{eq:fact}, such corrections arise from
power-suppressed interactions and production operators in SCET and PNRQCD and
require an extension of the factorization formula~\eqref{eq:fact} with
generalized hard, soft and potential functions.  In particular, the
simplification of the colour structure to that of a $2\to 1$ scattering
process has only been shown at leading power.  It is known that three-particle
colour correlations appear in infrared singular parts of the $pp'\to t\bar t$
scattering amplitudes~\cite{Ferroglia:2009ep,Ferroglia:2009ii}, but do not
contribute to the NNLO and NNLL cross
section~\cite{Beneke:2009ye,Beneke:2010da,Czakon:2013hxa}.
In the framework of~\cite{Beneke:2009rj,Beneke:2010da} these
corrections do not enter in the definition of the soft and hard functions and
their evolution equations, but
rather through the generalized soft and hard functions in the extended
factorization formula.  The complete
treatment of the $\mathcal{O}(\alpha_s^2,\alpha_s\beta, \beta^2)$ corrections
required for a full soft and Coulomb resummation at N$^3$LL is beyond the
scope of this paper, but we will include several corrections of this form:
  \begin{itemize}
  \item Corrections of the form $\alpha_s^3 \ln^{2,1}\beta$ to
    the N$^3$LO cross section arise from  (ultra)soft
    corrections\footnote{The ``soft'' modes in the
      terminology of~\cite{Beneke:2010da} are called ultrasoft in the PNRQCD
      literature.}   due to  the power-suppressed chromoelectric $\vec x\cdot
    \vec E_s$ vertex in the PNRQCD Lagrangian. In the language
    of~\cite{Beneke:2009rj,Beneke:2010da} these corrections arise from
    generalized soft and potential functions. We estimate these
    corrections using the known result for the
    colour-singlet case~\cite{Beneke:2008cr}.
  \item Corrections due to subleading production operators are related
    to the  P-wave production channels~\eqref{eq:pwave}, which 
    give rise  to $\alpha_s^3\times (\frac{1}{\beta}, \ln^2\beta,\ln\beta)$
    corrections relative to the leading S-wave channels since the $\beta^2$
    suppression of the Born cross section  can be compensated by the second Coulomb singularity.
    In
    Section~\ref{sec:pwave} we apply the NLL resummation formula to
    P-wave channels, which is sufficient to obtain the threshold-enhanced
    terms at $\mathcal{O}(\alpha_s^3)$.
  \item The interference of the second Coulomb correction with subleading
    soft-collinear interactions suppressed by $\beta^2$ results in
    contributions $\sim\alpha_s^3\log\beta$ to the total cross section.  While
    there is current interest in subleading soft-collinear effects using QCD
    factorization~\cite{Laenen:2010uz,Bonocore:2016awd,DelDuca:2017twk} and
    SCET~\cite{Larkoski:2014bxa,Moult:2017rpl,Feige:2017zci,Beneke:2017ztn} approaches, these
    effects have not been systematically studied for top-quark production.  In
    Section~\ref{sec:nte} leading-logarithmic subleading corrections to the
    initial state are heuristically included using universal results from
    Drell-Yan and Higgs production.
\end{itemize}
In this way, subleading soft-collinear and soft-potential corrections are
included.  From the arguments used to exclude contributions from power
suppressed soft-potential or soft-collinear interactions at
NNLO~\cite{Beneke:2010da} it is expected that sub-leading corrections at
N$^3$LO separately affect the initial and final state and a cross-talk between
soft-collinear and soft-potential corrections only appears at higher
orders.\footnote{Such a correction would involve one insertion of the
  chromoelectric vertex and one insertion of a subleading SCET interaction,
  where the latter vanish in a frame where initial-state partons have
  vanishing transverse momentum~\cite{Beneke:2010da,Larkoski:2014bxa}.}

\subsection{Hard functions}
\label{sec:hard}
The hard function satisfies  the evolution equation
\begin{equation}
\label{eq:rge-hard}
\frac{d}{d\ln\mu} H_{pp'}^{R,S}(\mu) =\left(
(\Gamma_{\text{cusp}}^{r}+\Gamma_{\text{cusp}}^{r'})\ln\left(\frac{4m_t^2}{\mu^2}\right)
+2\gamma^V_i  +\gamma_J^{R,S}(\mu) \right) H_{pp'}^{R,S}(\mu)  .
\end{equation}
Up to the three-loop level~\cite{Moch:2004pa}, the light-like cusp anomalous
dimension satisfies the so-called Casimir scaling,
i.e. $\Gamma_{\text{cusp}}^{r}=C_r \gamma_{\text{cusp}}$, where $C_r$ is the
quadratic Casimir for the $SU(3)$ representation $r$ of the parton $p$. At the
four-loop level, the recent result for the quark cusp-anomalous
dimension~\cite{Moch:2017uml} shows a violation of this property (see
also~\cite{Grozin:2017css,Boels:2017skl}).  The anomalous dimension
$\gamma^V_i$ can be written in terms of single-particle anomalous dimensions:
\begin{equation}
\label{eq:gamma-v}
\gamma^V_i=\gamma^p+\gamma^{p'}+\gamma_{H,s}^{R}.
\end{equation}
where the anomalous dimensions for light partons, $\gamma^p$, are known up to
three-loop level. The results in the notation used here are collected
in~\cite{Becher:2009qa}. The anomalous dimension for massive particles
$\gamma_{H,s}^{R}$ is known up to two-loop
level~\cite{Becher:2009kw,Beneke:2009rj}.  The anomalous dimension
$\gamma_J^{R,S}(\mu)$ arises first at two-loop order because of additional IR
divergences in the hard function, which are related to UV divergences of the
potential function arising from insertions of non-Coulomb potentials.  The RGE
for the corresponding matching coefficient of the electromagnetic quark
current was derived at NLL in the PNRQCD framework~\cite{Pineda:2001et} and
the fixed-order three-loop anomalous dimension was computed
in~\cite{Kniehl:2002yv}, see also~\cite{Hoang:2003ns}. For the colour-octet case, the
$\mathcal{O}(\alpha_s^2)$ result for $\gamma_J^{R,S}$ can be obtained from the
non-Coulomb corrections to the potential function obtained
in~\cite{Beneke:2009ye,Beneke:2016kvz}, and is given
in~\eqref{eq:gammaJ1}. The three-loop result is not known yet.

The hard functions are required up to the two-loop level for NNLL$'$ and
N$^3$LL resummation.  The one- and two-loop solution of the evolution
equation~\eqref{eq:rge-hard} is given explicitly in Appendix~\ref{app:h2}.
The initial conditions for the evolution can be obtained from comparing the
scaling functions obtained from the expansion of the resummation formula to
the threshold limit of the result of a diagrammatic calculation of the total
NLO and NNLO cross section in the corresponding colour and spin state.
Subtracting constant contributions to the potential and soft functions, this
relates the hard function at some initial scale to the constant (i.e.\
$\rho$-independent) term in the scaling function $f_{pp,(R)}^{(n,0)}$, which
will be denoted by $C_{pp}^{R(n)}$ (the spin-dependence of the scaling
functions will be left implicit for notational simplicity).  The resulting
relations of the NLO and NNLO constants $C_{pp}^{R(1)}$ and $C_{pp}^{R(2)}$ to
the one- and two-loop hard functions are given in~\ref{eq:c1} and~\ref{eq:c2}.
In order to extract the hard functions, we choose the initial scale
$\mu_h=m_t$, which simplifies the relation to the $C_{pp}^{R(n)}$ compared to
the more natural scale $2m_t$ since the factorization-scale dependence is
conventionally expressed in terms of $\ln(\frac{\mu_f}{m_t})$
in~\eqref{eq:scaling}.  The one-loop hard coefficients are obtained from the
comparison to the threshold expansion of the NLO cross
section~\cite{Hagiwara:2008df,Czakon:2008cx} as\footnote{In the expression for
  $h^{{\bf 8}(1)}_{gg}(\mu_h)$ in~\cite{Beneke:2011mq} the non-logarithmic
  terms with colour factor $C_A$ should have the opposite sign. The numerical
  implementation is not affected by this typo.}
\begin{align}
  h^{8(1)}_{q\bar q}(m_t)&=C_F\left(-8\ln^2 2+12\ln
    2-32+\frac{7\pi^2}{3}\right) 
  +\frac{C_A}{9}\left(-72\ln 2+200-9\pi^2\right)\nonumber\\
  &\qquad
  -\frac{4}{9}\left((10-12\ln 2) n_f T_F+8\right),\\
  h^{1(1)}_{gg}(m_t)&= C_F(-20+\pi^2)+\frac{4}{3}C_A(3+\pi^2-6\ln^2 2),\\
  h^{8_S(1)}_{gg}(m_t)&=C_F(-20+\pi^2)+C_A\left(8+\frac{5\pi^2}{6}-8\ln^2
    2+4\ln 2\right),
\end{align}
where the notation for the perturbative expansion of the hard function is
given in~\eqref{eq:hard-def}.

The results for the two-loop constants in the threshold limit have been
obtained in~\cite{Baernreuther:2013caa},
\begin{align}
  \bar C_{q\bar q}^{8(2)}&=1104.08-42.9666 n_f-4.28168 n_f^2,\\
  \bar C_{gg}^{1(2)}&=37.1457+17.2725 n_f,\\
  \bar C_{gg}^{8_s(2)}&=674.517-45.5875 n_f.
\label{eq:c28s}
\end{align}
We denote these constants by a bar to denote that the correction factors in the threshold expansion
in~\cite{Baernreuther:2013caa} are defined relative to the
leading cross sections~\eqref{eq:born} at $\hat s=4 m_t^2$, whereas we keep the
factor $4m_t^2/\hat s=1-\beta^2$ unexpanded. Because of the enhancement due to
the second Coulomb correction, this leads to a difference in the expression
for the two-loop constants in the two conventions, see Eq.~\eqref{eq:cbar}.
In addition, since no spin
decomposition is performed for the NNLO results~\cite{Baernreuther:2013caa},
 contributions from sub-leading P-wave production channels as given
  in~\eqref{eq:pwave-nnlo} below have to be subtracted.\footnote{After the
    completion of the work described here, results for polarized two-loop
    amplitudes have been published in~\cite{Chen:2017jvi}.}
A further contribution to the NNLO constant arises from the subleading
antisymmetric colour-octet gluon channel~\eqref{eq:born-8A} since the accidental suppression of
the Born matrix element does not persist at
one loop~\cite{Baernreuther:2013caa,Beenakker:2015mra,Cacciari:2015ela}.  Therefore, the one-loop
squared contribution in the $8_a$ channel leads to an
$\mathcal{O}(\alpha_s^2/\beta^2)$ correction relative to the leading order
cross section~\eqref{eq:born-8A}~\cite{Baernreuther:2013caa}
 \begin{equation}
\label{eq:c28a}
 \bar C_{gg}^{8_a(2)}=11.2531-2.29745 n_f+0.14257 n_f^2,
\end{equation}
which contributes at the same order as the
constants $C_{gg}^{R(2)}$ in the dominant colour-symmetric production
channels.
Since the top-quark pair is produced in an S-wave state in the $8_a$ channel and
the soft-gluon and Coulomb corrections do not distinguish between the
$8_s$ and $8_a$ representations, the
contribution to the resummed cross section can be correctly taken into account
by adding the constant~\eqref{eq:c28a}
 to the one of the
symmetric octet~\eqref{eq:c28s}. We therefore define
\begin{equation}
  \bar C_{gg}^{8(2)}=  \bar C_{gg}^{8_s(2)}+
  \frac{\sigma^{(0),1}_{gg}+\sigma^{(0),8_s}_{gg}}{\sigma^{(0),8_s}_{gg}}
  \bar C_{gg}^{8_a(2)}=690.271-48.8039 n_f+0.2 n_f^2,
\end{equation}
where the normalization of $C_{gg}^{8_a(2)}$ in eq.~(4.8)
of~\cite{Baernreuther:2013caa} was taken into account and $N_c=3$ was
used in the second step.
The constants $\bar C_{pp'}^{R(2)}$  are related to the two-loop coefficients of
the hard function by~\eqref{eq:h2-exp}. For
the number of light flavours $n_f=5$ and using $N_c=3$ and $T_F=1/2$ we obtain
\begin{align}
  h_{q\bar q}^{8(2)}(m_t)&=\bar C_{q\bar q}^{8(2)}-57.818=724.387,\\
  h_{gg}^{ 1(2)}(m_t)&=\bar C_{gg}^{1(2)}-1344.66=-1221.16,\\
  h_{gg}^{8(2)}(m_t)&=\bar C_{gg}^{8(2)}+143.403 =594.655.
\label{eq:h28s}
\end{align}

\subsection{Soft function}
\label{sec:soft}
The RGE of the Laplace-transformed  
soft function~\eqref{eq:soft-laplace} reads
\begin{equation}
\label{eq:rge-soft}
\frac{d}{d\ln\mu}  \tilde  s_i^{R}(\rho)
  =  \left[
    -(\Gamma_{\text{cusp}}^r+\Gamma_{\text{cusp}}^{r'})\, \rho 
    -2\gamma_{W,i}^R\right]
    \tilde s^{R}_i(\rho).
\end{equation}
The anomalous dimension $\gamma_W$ is related to that of the hard function
according to
\begin{equation}
\label{eq:gamma-w}
  \gamma^R_{W,i}=\gamma^V_i+\gamma^{\phi,p}+\gamma^{\phi,p'}
  =\gamma_s^r+\gamma_s^{r'}+\gamma_{H,s}^{R},
\end{equation}
with the anomalous dimensions $\gamma^{\phi,p}$ entering the evolution of the
parton distribution functions in the $x\to 1$ limit~\eqref{eq:splitting}.  In
the second equality of~\eqref{eq:gamma-w}, soft anomalous dimensions of
massless partons $p$ in the colour representation $r$, $\gamma_s^r\equiv
\gamma^p+\gamma^{\phi,p}$, have been introduced. At least up to three-loop
level for the massless partons and the two-loop level for massive particles,
the soft anomalous dimensions satisfy Casimir scaling
\begin{align}
  \gamma_s^r &= C_r \gamma_s, & 
  \gamma_{H,s}^R&=C_R\gamma_{H,s}.
\end{align}
The known results for the soft anomalous-dimension coefficients are summarized
in~\eqref{eq:gamma-s} and ~\eqref{eq:gamma-H}. The unknown three-loop soft
anomalous dimension for massive particles, $\gamma_{H,s}^{(2)}$, will be kept
explicitly in the results to estimate the resulting uncertainty.

The one- and two-loop solution of the evolution
equation~\eqref{eq:rge-soft} is given in Appendix~\ref{app:soft}. As initial
conditions, we require the functions $\tilde s_i^{R}(0)$,
which are available for arbitrary colour
representation~\cite{Beneke:2009rj} at one-loop level,
\begin{align}
  \tilde s_i^{(1)R}(0)&= (C_r+C_{r'})\frac{\pi^2}{6}+4C_R,
\end{align}
while the two-loop coefficients have been calculated for the colour
singlet~\cite{Belitsky:1998tc,Becher:2007ty} and octet~\cite{Czakon:2013hxa}
for identical initial state representations
(i.e. $C_{r}=C_{r'}$),
\begin{align}
  \tilde s_i^{(2)1}(0)&= C_r^2\frac{\pi^4}{18}
  +C_rC_A\left(\frac{2428}{81}+\frac{67\pi^2}{54}-\frac{\pi^4}{3}
    -\frac{22\zeta_3}{9}\right)\nonumber\\
  &\quad  +
  C_r n_f T_F
  \left(-\frac{656}{81}-\frac{10\pi^2}{27}+\frac{8\zeta_3}{9}\right),
 \nonumber \\
   \tilde s_i^{(2)8}(0)&= \tilde s_i^{(2) 1}(0)
   +C_A^2\left(\frac{1784}{27}+\frac{2\pi^2}{3}+\frac{13\pi^4}{180}-10\zeta_3
   \right)
   +C_A C_r \frac{4\pi^2}{3}- C_A n_f T_F\frac{640}{27}.
\label{eq:s20}
\end{align}
\subsection{Potential  corrections}
\label{sec:nnlo-pot}

The potential function for S-wave production in the colour channel
$R_\alpha$ and spin state $S$ is given in terms of the imaginary part of the
zero-distance Green function of the Schr\"odinger equation,
\begin{equation}  
J^{S}_{R_\alpha}(E,\mu)=2 \,\mbox{Im} \left[\, \left(1+d_J v^2\right) 
G_{R_\alpha}(0,0;E,\mu)  \right],
\label{JRal0}  
\end{equation}
where we use the abbreviation $v=\sqrt{E/m_t}$.  The LO Green function is
obtained as solution of the Schr\"odinger function with the Coulomb potential
and sums up all correction of the form $(\alpha_s/\beta)^k$. For higher-order
corrections we make use of the method described in~\cite{Beneke:2013jia} where
insertions of higher-order potentials are treated perturbatively. The N$^3$LO
Green function for the colour-singlet, spin-triplet case computed
in~\cite{Beneke:2013jia,Beneke:2015kwa,Beneke:2016kkb} includes all
corrections up to terms of order $(\alpha_s/\beta)^k\times
(\alpha_s^3,\alpha_s^2 v,\alpha_s v^2,v^3)$.

The coefficient  $d_J$ in~\eqref{JRal0} is introduced in order to reproduce the kinematical correction $(1-\frac{1}{3}\beta^2)$ to the S-wave production
processes~\eqref{eq:born}.
The expansion of the NNLO Green function 
reads $\mbox{Im} G_{R_\alpha}(0,0;E,\mu) \propto v(1+\frac{5}{8}
v^2)+\mathcal{O}(\alpha_s)$, where the kinematical correction $\sim v^2$
arises from the relativistic kinetic energy correction term $\partial^4/8m_t^3$ in the NRQCD Lagrangian.
Furthermore, the relation of the variable $v$ used in the PNRQCD Green function
to the non-relativistic velocity $\beta$ reads at this accuracy 
\begin{equation}
\label{eq:vbeta}
  v=\beta\left(1+\frac{3}{8}\beta^2+\dots \right).
\end{equation}
Taking these corrections into account leads to the value
\begin{equation}
  d_J=-\frac{4}{3}
\end{equation}
familiar from the treatment of top-quark pair production in $e^-e^+$ collisions.
The coefficient $d_J$ receives corrections at $\mathcal{O}(\alpha_s)$ and
becomes scale-dependent. However, these corrections only
contribute to the constant term at $\mathcal{O}(\alpha_s^3)$ and are therefore
beyond N$^3$LL accuracy.

Starting from NNLO the  potential function is
explicitly scale dependent,
\begin{equation}
\label{eq:rge-pot}
\frac{d}{d\ln\mu}  J_{R}^{S}(E,\mu)=-\gamma_J^{R,S}(\mu)  J_{R}^{S}(E,\mu),
\end{equation}
where the anomalous dimension is the same one that appears
in the evolution equation of the hard function~\eqref{eq:rge-hard}. 
In a fixed-order expansion to $\mathcal{O}(\alpha_s^3)$ it has the form
\begin{equation}
  \gamma_J^{R,S}(\mu)=\alpha_s^2 \gamma_{J}^{R,S(1)}
  +\frac{\alpha_s^3}{(4\pi)}
  \left[\gamma_{J}^{R,S(2,1)}\ln\left(\frac{2m_t}{\mu}\right)
    +\gamma_{J}^{R,S(2,0)}\right]+\mathcal{O}(\alpha_s^4).
\end{equation}
The three-loop anomalous-dimension coefficients $\gamma_{J}^{R,S(2,i)}$ are
known for the colour-singlet case~\cite{Kniehl:2002yv}.\footnote{The relation
  to the notation of~\cite{Kniehl:2002yv} is
  $\gamma_{J}^{R,S(1)}=2\gamma_v^{(2)}$, $\gamma_{J}^{R,S(2,1)}=-32
  \gamma_v^{(3)}$ and $\gamma_{J}^{R,S(2,0)}=8 (\frac{4}{3}(1+\ln
  2)\gamma_v^{(3)}+\gamma_v^{(3)'}-2\beta_0\gamma_v^{(2)})$. Note that the
  coefficient $\beta_0$ in the convention of~\cite{Kniehl:2002yv} is $1/4$
  that in our convention. A modification of $\gamma_v'^{(3)}$ due to the
  $d$-dimensional treatment of spin was pointed out
  in~\cite{Beneke:2007pj}.}
 The results for the colour-octet case are
currently not known.  However, in the fixed-order expansion of the cross
section to $\mathcal{O}(\alpha_s^3)$ these terms cancel against corresponding
contributions from the evolution of the hard function~\eqref{eq:rge-hard}.

\subsubsection{Potentials}
\label{sec:potentials}
For the computation of the Coulomb Green function we use the formalism of
PNRQCD in the conventions of~\cite{Beneke:2013jia}.
After performing a  colour and spin decomposition (see
Appendix~\ref{app:potential}), the  potential for a quark-antiquark pair can be written in the form
\begin{equation}
\label{eq:potential} 
V^{R,S}({\bf p},{\bf p}^{\prime}) =
  \frac{4\pi \alpha_{s}D_R}{{\bf q}^2} \bigg[\,{\cal V}^R_C
  -{\cal V}^R_{1/m}\,\frac{\pi^2\,|\bf{q}|} {m_t}
  +{\cal V}^{R,S}_{1/m^2}\,\frac{{\bf q}^2}{m_t^2}
  +{\cal V}_{p}^R\,\frac{{\bf p}^2+{\bf p}^{\prime \,2}}{2m_t^2}\,
\bigg]
+\frac{\pi \alpha_s}{m_t^2}\nua,
\end{equation}
with the colour factor 
\begin{equation}
  \label{eq:colour-coulomb}
D_R=\frac{C_R}{2}-C_F,   
\end{equation}
 where $C_R$ is the quadratic Casimir
of the representation $R$. 
For the computation of the $\mathcal{O}(\alpha_s^3)$ corrections to the
potential function, we require the Coulomb potential ${\cal V}^R_C$ and the
$1/m$ potential up to the two-loop level and the $1/m^2$ potentials ${\cal
  V}^{R,S}_{1/m^2}$ and ${\cal V}_{p}^R$ as well as the
so-called annihilation correction $\nua$ at one-loop. 
 The Coulomb potential at the relevant accuracy is
known both for the colour-singlet~\cite{Schroder:1998vy} and octet
states~\cite{Kniehl:2004rk} and is given in~\eqref{eq:coulomb}. 

The  $1/m^2$ potential is spin dependent and reads at leading order in
$d=4-2\epsilon$ dimensions 
\begin{equation}
{\cal V}^{R,S,(0)}_{1/m^2}
=\nus+\epsilon\,
\nu_{\text{spin}}^{\epsilon,S}+\mathcal{O}(\epsilon^2)
\end{equation}
 with the
explicit values
\begin{align}
  \nu_{\text{spin}}^{S=1}&=-\frac{2}{3} \,,&
    \nu_{\text{spin}}^{\epsilon,S=1}&=-\frac{5}{18} \,,\\
  \nu_{\text{spin}}^{S=0}&=0 \,,&
    \nu_{\text{spin}}^{\epsilon,S=0}&=-\frac{1}{2}.
\end{align}
The one-loop result for the colour-singlet case is given in~\cite{Beneke:2013jia}
and generalized to the octet case in Appendix~\ref{app:potential}, see also~\cite{Wuester:2003}. 

The annihilation contribution arises from local
four-fermion operators in NRQCD \cite{Pineda:1998kj} and was omitted
in the calculations of~\cite{Beneke:2009ye,Beneke:2011mq}.
The leading  correction arises for the
colour-octet, spin-triplet case,
\begin{equation}
    {\nu_{\mathrm{ann}}^{8,1}}^{(0)}=1,
\end{equation}
and results in a contribution to the  NNLO threshold expansion of the
top-quark pair production cross section in the quark-antiquark channel
first obtained by an explicit two-loop calculation
in~\cite{Baernreuther:2013caa}.
The annihilation corrections to the one-loop matching coefficients of
the four-fermion operators~\cite{Pineda:1998kj}  also arise in the 
spin-singlet channel and therefore also enter in the gluon-fusion
initial state. The corresponding one-loop coefficients and the
resulting corrections to the potential function are given in Appendix~\ref{app:annihilation}.

Therefore, all ingredients of the potential are known at the required
accuracy, apart from  the $1/m$ potential, which is known at the one-loop level~\cite{Beneke:2009ye,Beneke:2016kvz}  for the singlet and
octet cases, but only for the singlet case at the two-loop level~\cite{Kniehl:2001ju}.  In
addition, at $\mathcal{O}(\alpha_s^3)$ also ultrasoft corrections due to the
power-suppressed $\vec x\cdot \vec E_s$ vertex in the PNRQCD Lagrangian become
relevant, which are also only known for the colour-singlet
case~\cite{Beneke:2008cr}.  We will estimate the effect of the $1/m$ potential
and the ultrasoft corrections in the colour-octet case by a naive replacement
of colour factors.

\subsubsection{Expansion of the potential function}

Given the above results for the potentials, we can obtain the
expansion of the potential function to ${\cal O}(\alpha_s^3)$ using
the expressions from the N$^3$LO calculation of $e^-e^+\to
t\bar{t}$~\cite{Beneke:2013jia,Beneke:2015kwa,Beneke:2016kkb}. The annihilation corrections are computed as discussed in Appendix~\ref{app:annihilation}.
A brief
description of the methods used for the expansion of the Green
function in $\alpha_s$ is given in the following. The
required corrections 
are expressed in terms of
nested harmonic sums and sums over (poly-)gamma functions. They depend
on $\alpha_s$ through the parameter
$\lambda=(-D_R\alpha_s/2)\sqrt{-m_t/(E+i\Gamma_t)}$, where $\Gamma_t$
is the top-quark decay width. In the former case, the summation limits
depend on $\lambda$, but we can always transform the sums in such a
way that this dependence is shifted to the summand. In both cases, we
then expand the summand in the limit $\lambda\to 0$ before performing
the sum. Afterwards, the coefficients of the expansion are given by
so-called multiple zeta values, which we take from the program
{\tt summer}~\cite{Vermaseren:1998uu}. A simple example for the series
expansion of a single harmonic sum is
\begin{eqnarray}
  S_a(-1-\lambda) &=& \sum\limits_{i=1}^{-1-\lambda} \frac{1}{i^a} =
  S_a(\infty) - \sum\limits_{i=-\lambda}^{\infty} \frac{1}{i^a} =
  S_a(\infty) - \sum\limits_{i=0}^{\infty} \frac{1}{(i-\lambda)^a}
                      \nonumber\\
  &=& S_a(\infty) - \left[ \frac{1}{(-\lambda)^a} +
      \sum\limits_{i=1}^{\infty} \frac{1}{i^a}
      \sum\limits_{n=0}^{\infty} \frac{\Gamma(a+n)}{\Gamma(a)\, n!}
      \left(\frac{\lambda}{i}\right)^n \right] \nonumber\\
  &=& -\frac{1}{(-\lambda)^a} - \sum\limits_{n=1}^{\infty}
      \zeta(a+n)\, \frac{\Gamma(a+n)}{\Gamma(a)\, n!}\, \lambda^n \,,
\end{eqnarray}
where $\zeta(n)=S_n(\infty)$ is Riemann's zeta function with integer
argument. Since polygamma functions are related to harmonic sums,
after expanding in $\alpha_s$ we obtain similar expressions in that
case as well. The infinite series can then be truncated at the
required order in $\alpha_s$. Finally, we can take the limit
$\Gamma_t\to 0_+$ (with $E>0$), which yields the imaginary part of the
Green function from the discontinuity of the square root in
$\lambda$. 
Bound-state effects arising for $E<0$ as
 taken into account in the resummed calculation~\cite{Beneke:2011mq} are not
 included in this naive expansion. 
In the $\mathcal{O}(\alpha_s^3)$ corrections, these bound-state
effects give rise to a contribution localized at $E=0$~\cite{Beneke:2016jpx}, which is added
separately below.

The resulting expansion of the potential function in the strong coupling
constant is written as
\begin{equation}
\label{eq:pot-exp}
J_R^{S}(E,\mu)=2 \frac{m_t^2}{(4\pi)} v
\left(1+d_J v^2\right)
\left[1+\frac{5}{8}v^2 -\frac{\pi D_R\alpha_s(\mu)}{2} \frac{1}{v}+ \sum_{n=2}^\infty \alpha_s^n(\mu) \Delta J_{R}^{S(n)}(E,\mu)\right]\theta(E).
\end{equation}
The $\mathcal{O}(\alpha_s^2)$ corrections to the
potential function read
\begin{align}
 \Delta J_{R}^{S(2)}(E,\mu)&= 
\frac{\pi ^2 D_R^2}{12}\frac{1}{v^2}
+\frac{1}{v}
\frac{D_R}{8}\left( -2\beta_0 L_E-a_1\right)
- \gamma_J^{R,S(1)} L_E + c_{J,2},
\label{eq:JNNLO2}
\end{align}
where  the scale dependence enters only
through the variable
\begin{align}
  L_E=-\ln\left(\frac{2 v m_t}{\mu}\right).
\end{align}
The constant $a_1$ arises from the NLO  Coulomb potential and is given
in~\eqref{eq:a1}.
The $\mathcal{O}(\alpha_s^2)$ coefficient of the anomalous
dimension $\gamma_J^{R,S}$ entering the RGE~\eqref{eq:rge-pot} can be
read off the expansion of the potential function and is given by 
\begin{equation}
  \gamma_J^{R,S(1)}=
  D_R \left(C_A-2D_R\left(\nus+1\right)-\frac{{\nua}^{(0)}}{2}\right).
\label{eq:gammaJ1}
\end{equation}
For ${\nua}^{(0)}=0$ and the colour-singlet case $D_1=-C_F$ this result agrees with the 
NLL anomalous dimension of the electromagnetic quark current in eq.~(14)
of~\cite{Pineda:2001et} for the fixed-order values of the potential
coefficients $D_s^{(1)}=\alpha_s^2(\mu)$ and $D^{(2)}_{S^2,s}=D^{(2)}_{d,s}=D^{(2)}_{1,s}=\alpha_s(\mu)$, as well as with~\cite{Kniehl:2002yv}.
The constant term in~\eqref{eq:JNNLO2} is given by
\begin{equation}
\label{eq:cj2}
  c_{J,2} 
=D_R^2\left(\frac{\nuseps}{2}+\frac{9 \pi^2}{32}+\frac{9}{4} \right)
-\frac{3}{4}C_A D_R- \frac{1}{2}\gamma_J^{R,S(1)}.
\end{equation}
All terms in~\eqref{eq:JNNLO2} apart from the constant were obtained
in~\cite{Beneke:2009ye}, with the exception of the annihilation contribution
which was added in~\cite{Baernreuther:2013caa,Beneke:2016kvz}.

The computation of the  $\mathcal{O}(\alpha_s^3)$ corrections to the potential
function requires the use of
the N$^3$LO Green function, which includes all corrections of the
form $\alpha_s^3\times (\alpha_s/\beta)^k$.
The result can be split into several contributions:
\begin{equation}
  \label{eq:jas3}
  \Delta J_{R}^{S(3)}(E)=  \Delta J_{R,\text{LO}}^{(3)}(E)+ 
  \Delta J_{R,\text{NNLO}}^{S(3)}(E)+ 
  \Delta J_{R,\text{N}^3\text{LO}}^{S(3)}(E).
\end{equation}
The expansion of the LO-Green
function includes a term $\sim \alpha_s^3/E$ whose imaginary part is nonvanishing for unstable particles (see e.g.~\cite{Actis:2008rb}) but does not yield a correction of the form $\alpha_s^3/v^3$ if the decay width is neglected.
Treating the imaginary part of the Green function carefully in the
distributional sense, it was shown that there is instead a delta-function contribution at $\mathcal{O}(\alpha_s^3)$~\cite{Beneke:2016jpx}: 
\begin{equation}
   \Delta J_{R,\text{LO}}^{(3)}(E)=- \alpha_s^3D_R^3 \frac{m_t^3}{8}\zeta_3 \delta(E).
\label{eq:c3dist}
\end{equation}
As discussed in~\cite{Beneke:2016jpx} this contribution is taken into
account in the resummed
calculation~\cite{Beneke:2011mq,Beneke:2012wb}, which includes the
resummed Coulomb corrections above threshold and the bound-state
corrections below.

The naive expansion
of the NNLO Green function in $\alpha_s$ is fully known for all
required colour and spin states and gives contributions
enhanced by one inverse power of $\beta$. The scale-dependence again enters
only through the variable $L_E$:
\begin{align}
  \label{eq:jnnlo3}
\Delta J_{R,\text{NNLO}}^{S(3)}(E)&=\frac{1}{4\pi} \Biggl\{
\frac{1}{v^2}\frac{D_R^2}{6}\left[
    \pi^2  \left( 2\beta_0 L_E +a_1\right)
         -12\beta_0\zeta_3\right] \nonumber\\
&\quad +  \frac{1}{v}
D_R\Biggl[
-\frac{1}{2}\beta_0^2 L_E^2 
+\frac{1}{8} \left((4\pi)^2\gamma_J^{R,S(1)}-2\beta_1-4a_1\beta_0\right)L_E
+c_{J,3}^{(0,1)}  \Biggr] \Biggr\},
\end{align}
with the constant
\begin{equation}
c_{J,3}^{(0,1)}=
-\frac{a_2^R}{8}+\pi^2\gamma_{J}^{R,S(1)}- \frac{\beta_0^2\pi^2}{12}
  +\frac{3\pi^2}{2} D_R C_A -\pi^2 D_R^2\left(\nuseps+\frac{9}{2}\right),
\end{equation}
where the coefficient $a_2^R$ of the NNLO Coulomb potential is given
in~\eqref{eq:a2}.  All contributions of~\eqref{eq:jnnlo3} apart from the
constant $c_{J,3}^{(0,1)}$ are included in the implementation of NNLL resummation
in~\cite{Beneke:2011mq} and the corresponding approximate N${}^3$LO
prediction.  For the colour-singlet case and the spin states $S=1$ and $S=0$,
this result reproduces the imaginary part of the threshold expansion of the
vector and pseudo-scalar current correlators given in the Appendix
of~\cite{Kiyo:2009gb} if the corresponding matching
coefficients~\cite{Czarnecki:1997vz,Beneke:1997jm,Kniehl:2006qw} are taken
into account.\footnote{In the pseudo-scalar case the so-called singlet
  contributions to the two-loop matching coefficient~\cite{Kniehl:2006qw} have
  to be set to zero since they are not included in~\cite{Kiyo:2009gb}.}

The expansion of the N$^3$LO correction to the Green function to
$\mathcal{O}(\alpha_s^3)$ gives rise to purely logarithmic
corrections~$\alpha_s^3\ln^{2,1}\beta$ and scale-dependent terms
$\alpha_s^3\ln^{2,1}(\frac{m_t}{\mu})$ governed by the RGE~\eqref{eq:rge-pot}.
 The result can be written in the form
\begin{align}
  \label{eq:jn3lo}
 \Delta J_{R,\text{N}^3\text{LO}}^{S(3)}(E)&=\frac{1}{4\pi} \Biggl\{
\left(
  c_{J,3}^{(2,0)}-2\beta_0 \gamma_J^{R,S(1)}\right) \ln^2 v 
+ c_{J,3}^{(1,0)}\ln v\nonumber\\
&\qquad 
-4\beta_0\gamma_J^{R,S(1)}\ln v \ln\left(\frac{2m_t}{\mu}\right) 
+\frac{1}{2}\left(\gamma_{J}^{R,S(2,1)}
    -4\beta_0\gamma_J^{R,S(1)}\right)
  \ln^2\left(\frac{2m_t}{\mu}\right)  \nonumber\\
&\qquad 
+\left( -4\beta_0  c_{J,2} 
+\gamma_{J}^{R,S(2,0)}\right)   \ln\left(\frac{2m_t}{\mu}\right)
\Biggr\}.
\end{align}

The result of~\cite{Beneke:2013jia,Beneke:2015kwa,Beneke:2016kkb} and the
generalization of the $1/m^2$ potential to general spin  in~\eqref{eq:oneoverm2} allows to compute the coefficients
$c_{J,3}^{(i,0)}$ and the anomalous-dimension coefficients
$\gamma_{J}^{R,S(2,i)}$ exactly for the colour-singlet state.  For the
colour-octet case, we use the result for the $1/m^2$ potential~\eqref{eq:oneoverm2} and estimate the unknown $1/m$ potential and the
ultrasoft corrections by a naive replacement of colour factors.  The ultrasoft
corrections to the Green function in~\cite{Beneke:2008cr} are presented in a
form where logarithmic terms $\ln\alpha_s$, $\ln\frac{\mu}{m_t}$ are given
explicitly while a function $\delta^{\text{us}}(\hat E)$ of the variable $\hat
E=\frac{E}{m_t\alpha_s^2}$ is only available numerically.  Using the fact that
$\ln\alpha_s$-terms cannot appear in a fixed-order calculation and thus must
be canceled by $\log\hat E$-terms in the remainder function
$\delta^{\text{us}}(\hat E)$, it can be checked that all logarithmic
$\mathcal{O}(\alpha_s^3)$-contributions can be reconstructed with the replacement
$\ln\alpha_s\to \ln v$.

We have checked that the expansion of the Green function computed
in~\cite{Beneke:2013jia,Beneke:2015kwa,Beneke:2016kkb} reproduces the
three-loop anomalous-dimension coefficients $\gamma_{J}^{R,S(2,i)}$ for the
colour-singlet case~\cite{Kniehl:2002yv}.\footnote{In this comparison the
  scale-dependence of the coefficient $d_J$ in~\eqref{JRal0} must be
  taken into account, see e.g. Eq.~(3.45) in~\cite{Beneke:2013jia} for the case of
  the vector current.}  The result for the remaining
coefficients is given by
\begin{align}
 c_{J,3}^{(2,0)} & =-D_R\left(\frac{14
    D_R^2}{3}+(7\nus+3) C_A D_R+
 \frac{1}{2}\beta_0 \,{\nua}^{(0)}\right) 
  +\delta c_{J,3}^{(2,0)},
    \\
 c_{J,3}^{(1,0)}  &=- D_R ^3\left(\frac{38}{3}+4\ln 2 +12\nus  \right) 
+D_R^2\Biggl[ - C_A\left(\frac{197}{9}+\frac{8 \pi ^2}{9}-\frac{34}{3} \ln 2\right) \nonumber\\
&\quad
 -4C_F\left(1+\frac{4}{3}\ln 2\right) +\nus\left( -8 C_F +\frac{40}{9}n_l T_F+2\beta_0-C_A
  \left(\frac{107}{9}+14\ln 2 \right) \right) \nonumber\\
&\quad 
+T_F\left(\frac{40}{9} n_l-\frac{8}{15}\right)
\Biggr]
+ D_R\beta_0 \left(4C_A-D_R \left( 8+\frac{9\pi^2}{8}\ \right)\right) 
\nonumber\\
&\quad +a_1\left(-D_R(C_A+ D_R)+\gamma_J^{R,S(1)} \right)
-\frac{D_R}{2}{\nua}^{(1)}(2m_t)+ \delta c_{J,3}^{(1,0)}.
\end{align}
The terms given explicitly in these expressions hold both in the colour-singlet and octet cases. 
The contributions from the $1/m$ potential and
ultrasoft corrections in the colour-singlet case are given by
\begin{subequations}
\label{eq:unknown-pot}  
\begin{align}
  \delta c_{J,3}^{(2,0)}&=\frac{4C_F}{3}\left(4 C_A^2+8 C_A
    C_F\right)-C_F \frac{4760}{27},\\
  \delta c_{J,3}^{(1,0)}&=-\frac{1}{9} C_A^2 C_F
(48 \ln 2+197)-\frac{2}{9} C_A C_F^2 (48 \ln 2+31)
+n_l T_F \frac{1}{9} (49 C_A C_F-8 C_F^2)\nonumber\\
&\qquad 
-\frac{16}{81} C_F \left(969 \ln 2-52 \pi ^2-1331\right),
\end{align}
\end{subequations}
where the terms containing only the single colour factor $C_F$ arise from the
ultrasoft corrections.\footnote{Note that in the calculation of the ultrasoft
  corrections~\cite{Beneke:2008cr} the explicit colour factors for $N_c=3$
  were used.} 

\subsection{P-wave contributions}
\label{sec:pwave}

The gluon-fusion initial state to top-quark pair production
receives contribution from the ${}^3\mathrm{P}_0$ and  ${}^3\mathrm{P}_2$
production channels~\eqref{eq:pwave}, which contribute at the same order in
the threshold expansion as the kinematic correction to the leading S-wave
production channels taken into account in~\eqref{JRal0}.
The relative P-wave contribution to the total cross section is parameterized by
the quantity
\begin{equation}
h^{(0),R}_{pp',P} 
=\frac{\sum_J\sigma^{(0),R,{}^S\mathrm{P}_J}_{pp'}}{\beta^2\sigma^{(0),R}_{pp'}},
\end{equation}
where $\sigma^{(0),R}_{pp'}$ is the threshold limit of the leading S-wave
production cross section~\eqref{eq:born}.
For top-quark pair production, P-wave production contributes only to the gluon
initial state in the symmetric colour representations with
\begin{equation}
\label{eq:hP}
  h^{(0),R}_{gg,P}=\frac{7}{3}\;, \quad R=1,8_s.
\end{equation}

The leading-order P-wave potential function for
stable particles above threshold is given by~\cite{Bigi:1991mi,Beneke:2013kia}
\begin{equation}
\label{eq:P-GF}
\begin{aligned}
  J_{R_\alpha}^P (E)&= m_t^2 v^2 \left(1+\frac{(\alpha_sD_{R_\alpha})^2 }{4v^2}
 \right)  J_{R_\alpha} (E)\,,\qquad E>0,
\end{aligned}
\end{equation}
with the perturbative expansion
\begin{equation}
    J_{R}^P (E)= \frac{m_t^4}{2\pi} v^3 \left[
  1 - \frac{\pi D_R\alpha_s}{2} \frac{1}{v}
    +\frac{\alpha_s^2
      D_R^2(3+\pi^2)}{12}\frac{1}{v^2}
    - \frac{\alpha_s^3 \pi D_R^3}{8} \frac{1}{v^3}\dots
\right].
\label{eq:P-GF-exp}
\end{equation}
 The $\alpha_s^3$ correction agrees with explicit results of the
 threshold expansion of the current
correlation function~\cite{Penin:1998ik,Kiyo:2009gb}.

The combination of the Born suppression $\sim \beta^2$ of the P-wave cross
section and the two-loop Coulomb enhancement leads to a constant $\alpha_s^2$
contribution relative to the leading S-wave production channel and
threshold-enhanced N$^3$LO contributions of order $\alpha_s^3/\beta$ and
$\alpha_s^3\ln\beta$.  In order to obtain these corrections, it is sufficient
to consider soft-gluon resummation at NLL accuracy.  In~\cite{Falgari:2012hx}
the factorization~\eqref{eq:fact} has been shown at least at NLL for partonic
channels with leading P-wave production. Here we heuristically apply this
result also for the case of subleading P-wave contributions to an S-wave
dominated process.\footnote{Complications may arise as in the related case of
  QCD calculations for quarkonia, where P-wave and S-wave processes are mixed
  by radiative corrections. However, at least at NLO these issues arise for
  the decay of quarkonia to quark pairs, but not in quarkonium production from
  gluon fusion~\cite{Petrelli:1997ge}.}  We therefore insert the P-wave
  potential function in the factorization formula~\eqref{eq:fact} and use the
  resummed soft and hard functions at NLL accuracy, setting the soft- and hard
  scales appearing in the momentum-space resummation formalism to $\mu_s= m_t
  \ln^2\beta$ and $\mu_h= m_t$. Leading logarithms of potential origin are
  taken into account by using the scale $\mu_C= m_t\ln\beta$ in the potential
  function.

The constant contribution at NNLO obtained in this way is given by
\begin{equation}
\label{eq:pwave-nnlo}
  \Delta\sigma_{pp', P}^{R(2)}=
 \sigma_{pp'}^{(0),R} \frac{\alpha_s^2 D_R^2\left(3+\pi^2\right)}{12}
h^{(0),R}_{pp',P}.
\end{equation}
This contribution was taken into account in the determination of the
two-loop hard function in Section~\ref{sec:hard}.
The relevant threshold-enhanced corrections
to the N$^3$LO cross section are obtained as
\begin{equation}
\label{eq:pwave-n3lo}
  \Delta \sigma_{pp', P}^{R(3)}
=\sigma_{pp'}^{(0),R}%
h^{(0),R}_{pp',P}  \frac{\alpha_s^3}{(4\pi)^3}  
\sum_{m=0}^1f^{(3,m)P}_{pp'(R)} \ln^m\left(\frac{\mu_f}{m_t}\right),
\end{equation}
with the scaling functions
\begin{align}  
f^{(3,0)P}_{pp'(R)}
&= -\frac{8 \pi^4 D_{R}^3}{\beta} 
+  \frac{4\pi^2  D_R^2}{3}\left(3+\pi^2\right)\Bigl\{16 (C_r+C_{r'}) \ln^2\beta
 \nonumber\\
& \quad -4\left[ 2(C_{R}+ (C_r+C_{r'})(4-6\ln 2)\right)+\beta_0]\log\beta
\Bigr\}+\mathcal{O}(1),\\
f^{(3,1)P}_{pp'(R)}
&=- \frac{4\pi^2  D_R^2}{3}\left(3+\pi^2\right)16 (C_r+C_{r'}) \ln\beta +\mathcal{O}(1).
\end{align}  
Note that the form  of these corrections relative to the ``constant'' NNLO
term~\eqref{eq:pwave-nnlo} differs from the usual NLO threshold corrections
for an S-wave production process (see e.g.~\cite{Beneke:2010da}) by the
coefficient of the Coulomb correction and the presence of the
 $\beta_0$-term that arises from the LL-running of the
potential function.

\subsection{Next-to-eikonal correction}
\label{sec:nte}
Contributions of the order $\alpha_s^3\ln\beta$ can arise from the
interplay of 
power-suppressed NLO corrections $\alpha_s \ln\beta\times\beta^2$ and the NNLO
Coulomb correction $\alpha_s^2/\beta^2$. A full analysis of these corrections
in the EFT framework is beyond the scope of this paper.
However, for the N$^3$LO threshold approximation only the LL next-to-eikonal
corrections are relevant that arise only from initial-state radiation and
therefore can be obtained from results for the Drell-Yan
process and Higgs production (see e.g.~\cite{Kramer:1996iq,Laenen:2010uz,DelDuca:2017twk}).

These corrections can be incorporated in our framework by including an
additional term in the one-loop soft function obtained by the replacement
\begin{equation}
  \left[\frac{\ln(1-x)}{1-x}\right]_+\to
  \left[\frac{\ln(1-x)}{1-x}\right]_+-
  \ln(1-x)
\end{equation}
 in the
position-space soft function $W_i^{R}(2m_t(1-z))$ given in eq. (C.5)
in~\cite{Beneke:2009rj}. This leads to the LL next-to-eikonal contribution 
\begin{equation}
\label{eq:ne-soft}
  \Delta W_{i,\text{n.e.}}^{(1)}(2m_t(1-z))=-\frac{1}{2 m_t}8(C_r+C_{r'})\ln(1-z),
\end{equation}
where the factor $(2m_t)^{-1}$ arises from the normalization convention of the
LO soft function, $W_i^{(0)R}(\omega)=\delta(\omega)$.

The next-to-eikonal correction to the cross section is obtained by
inserting~\eqref{eq:ne-soft} into the factorization
formula~\eqref{eq:fact}\footnote{Note
  that we do not claim that this factorized formula holds to all orders in
  perturbation theory and beyond the LL next-to-eikonal level.}
\begin{equation}
 \Delta \hat\sigma_{pp',\text{n.e.}}(\hat s,\mu)
= \sum_{R={\bf 1},{\bf 8}}H^{R,S}_{pp'}(m_t,\mu)
\;\frac{\alpha_s(\mu)}{(4\pi)}\;\int_\rho^1 dw\;
J_{R}^S(m_t(w-\rho))\,
 2m_t\Delta W_{i,\text{n.e.}}^{(1)}(2m_t(1-w))\,,
\end{equation}
where we have performed a transformation of variables $\omega=2m_t(1-w)$ and
approximated $E=\sqrt{\hat s}-2m_t\approx m_t(1-\rho)$.
The only threshold-enhanced contribution at N$^3$LO arises from the second
Coulomb correction in the NNLO potential function~\eqref{eq:JNNLO2}.  We
therefore obtain the logarithmic correction to the N$^{3}$LO scaling function
as
\begin{equation}
  \label{eq:ne-scaling}
  \Delta_{\text{n.e.}} f_{pp'(R)}^{(3,0)} =-(4\pi^2)^2 D_R^2 (C_r+C_{r'})\frac{8}{3}\ln\beta  .
\end{equation}

\section{Approximate N$^3$LO results}
\label{sec:results}
We are now in a position to obtain approximate N$^3$LO predictions based on
the partial N$^3$LL resummation  constructed in the previous section.
In Section~\ref{sec:partonic} we present explicit results for the scaling
functions and discuss our estimate of the currently unknown colour-octet
coefficients. In Section~\ref{sec:hadronic} we present results for the total
hadronic top-quark pair production cross section and discuss the scale
dependence and the estimate of the remaining theoretical uncertainty.

\subsection{Partonic cross sections}
\label{sec:partonic}

Our default implementation of the partial N$^3$LL resummed cross section is
obtained from the resummation formula~\eqref{eq:resum-sigma} by inserting the
N$^3$LO potential function, the resummation functions
$U_i$ and $\eta$ at N$^3$LL accuracy, and the two-loop hard and soft
functions obtained in Sections~\ref{sec:hard} and~\ref{sec:soft}. In addition,
the  P-wave contribution~\eqref{eq:pwave-n3lo} and the contribution from the
interference of next-to-eikonal logarithms and the Coulomb
corrections~\eqref{eq:ne-scaling} are added.
The resummation formula is expanded to $\mathcal{O}(\alpha_s^3$), setting the soft and hard scales
introduced in the resummation formalism to $\mu_s=k_s m_t v^2$ and
$\mu_h=k_h m_t$, where the unknown constants $k_X$ are of order one. For
N$^3$LL resummation, all threshold-enhanced N${}^{3}$LO terms are independent
of these unphysical parameters. 

We will present the results for the approximate N$^3$LO corrections for the
different partonic production channels in the
form 
\begin{equation}
\Delta \hat \sigma_{pp'}^{R(3)}(\hat s,m_t,\mu_f) =
\hat \sigma^{R(0)}_{pp'}(\hat s,m_t,\mu_f) 
\left(\frac{\alpha_s(\mu_f)}{4\pi}\right)^3
 \sum_{m=0}^3 f^{(3,m)}_{pp'(R)}(\rho) \ln^m\left(\frac{\mu_f}{m_t}\right),
\label{eq:sigma3-series}
\end{equation}
where the threshold limit of the partonic production
cross sections for a given colour channel~\eqref{eq:born} was factored out. The
renormalization-scale dependence can be obtained using~\eqref{eq:mur}.
We do not perform a decomposition into states of definite spin, i.e. the
scaling functions for the gluon initial states include the P-wave
contribution~\eqref{eq:pwave-n3lo}.

The result of the $\omega$-integral over the potential function in the
resummation formula~\eqref{eq:resum-sigma} is expressed in terms of the energy
variable $E=\sqrt{\hat s}-2m_t\equiv m_t v^2$.  To write the approximate
N$^3$LO corrections as functions of
the customary variable $\beta=\sqrt{1-\rho}$, the terms $\sim \beta^2$ in the
relation of the two variables~\eqref{eq:vbeta} must be included.
 Taking the
overall factor $v$ in the potential function~\eqref{eq:pot-exp} into account,
the relevant corrections to the threshold-enhanced terms at $\mathcal{O}(\alpha_s^3)$ can be obtained by the replacements
\begin{align}
 \frac{1}{v^2}&\to\frac{1}{\beta^2}-\frac{3}{8},&
  \frac{\ln v}{v^2}&\to \frac{\ln\beta}{\beta^2}
  -\frac{3}{8} (\ln\beta-1), &
 \frac{\ln^2 v}{v^2}&\to \frac{\ln^2\beta}{\beta^2}
  -\frac{3}{8}\left(\ln^2\beta-2\ln\beta\right),
\end{align}
where the $\mathcal{O}(1)$ terms contribute to the constants $C_{pp'}^{(3)}$ in the
scaling functions. 
Note that  in contrast to the NNLL implementation~\cite{Beneke:2011mq}, at this order one should not replace the threshold limit 
$\sigma^{R(0)}_{pp'}$ in~\eqref{eq:sigma3-series} by the full Born cross
section, since this
would treat these kinematic corrections in an inconsistent way.
In order to estimate kinematic ambiguities of the threshold approximation, we
will also consider the results obtained with the un-expanded scaling functions
in terms of the variable $v$.

The results for threshold expansion of the
scale-independent scaling functions for the different production channels are
for the explicit values $N_c=3$, $n_f=5$, 
\begin{subequations}
\label{eq:N3LO}
\begin{align}
f^{(3,0)}_{q\bar q(8)}=&
12945.4 \ln^6\beta-37369.1
   \ln^5\beta+27721.4 \ln^4\beta+41558.7 \ln^3\beta
\nonumber\\
  & +(-32785.3 + 157.914 \,\delta c_{J,3}^{(2,0)})
   \ln^2\beta
  \nonumber\\
  & +(2611.05 +157.914\, \delta c_{J,3}^{(1,0)}+12\gamma_{H,s}^{(2)}) \ln \beta  \nonumber\\
  &
  +\frac{1}{\beta}\Bigl[-2994.51 \ln^4\beta +2804.73 \ln^3\beta+    3862.46 \ln^2\beta - 6506.96 \ln\beta-2774.26\Bigr]
\nonumber\\
&+\frac{1}{\beta^2} \left[153.93 \ln^2\beta+122.866 \ln \beta-144.996\right]
 + C^{(3)}_{q\bar q(8)}\, ,\\[.2cm]
f^{(3,0)}_{gg(1)}=& 147456.0 \ln^6\beta-59065.60 \ln^5\beta-286099.5 \ln^4\beta+349462.5 \ln^3\beta
\nonumber\\
&-117661.0 \ln^2\beta
 -117672.0 \ln\beta 
+\frac{1}{\beta}\Bigl[121277.7 \ln^4\beta \nonumber\\
&
+103557.4 \ln^3\beta-164943.8 \ln^2\beta+56418.52 \ln\beta +14838.1 \Bigr]
\nonumber\\
&
+\frac{1}{\beta^2}\left[22165.98 \ln^2\beta+39012.06\ln\beta-2876.606\right]
+ C^{(3)}_{gg(1)}\, ,\\
f^{(3,0)}_{gg(8)}=& 147456.0 \ln^6\beta-169657.6 \ln^5\beta
   -140833.9 \ln^4\beta +524210.4 \ln^3\beta \nonumber\\
  &+(-298530.0 +157.9137\,\delta c_{J,3}^{(2,0)}) \ln ^2\beta \nonumber\\
  &+
  (48175.5+12 \gamma_{H,s}^{(2)} +157.9137\,\delta c_{J,3}^{(1,0)})\ln\beta\nonumber\\
  &
 +\frac{1}{\beta}  \Bigl[-15159.71 \ln^4\beta-5364.824 \ln^3\beta +19598.89 \ln^2\beta-17054.74 \ln\beta-2775.05\Bigr]  \nonumber\\
  &+\frac{1}{\beta^2}\left[346.3434 \ln^2\beta+522.9776 \ln\beta-71.78836\right]
 + C^{(3)}_{gg(8)}\, .
\end{align}
\end{subequations}
The coefficients of the terms $\ln^{6,\dots 3}\beta$,
$\frac{1}{\beta}\ln^{4,\dots 1}\beta$ and $\frac{1}{\beta^2}\ln^{2,\dots 0}\beta$
reproduce the results of the expansion of the NNLL cross section
in~\cite{Beneke:2011mq} (up to shifts $41839.4\to 41558.7$ in the coefficient
of the $\ln^3\beta$ term and $ 6528.61\to 6506.96$ in the $\ln\beta/\beta$
term in the quark-antiquark channel due to the annihilation contribution,
which was omitted in~\cite{Beneke:2011mq}).
The threshold approximation of the scaling functions $f^{(3,i)}$ for $i\neq 0$
related to the factorization-scale dependence are given in Appendix~\ref{app:scaling}.
These results were obtained both
from the expansion of the resummation formula and by exploiting the known
factorization scale dependence of the PDFs as discussed in
Section~\ref{sec:setup}, finding agreement with the results obtained using the
two methods. 

The results for the colour-octet channels in~\eqref{eq:N3LO} depend on the
unknown three-loop coefficient of the massive soft anomalous dimension
$\gamma_{H,s}^{(2)}$ and the coefficients $\delta c_{J,3}$ parameterizing the
unknown corrections to the potential function from the $1/m$ potential and
ultrasoft corrections. 
In our default predictions we set the three-loop massive soft anomalous dimension to zero
and estimate the uncertainty based on the known one- and two-loop results: 
\begin{equation}
\label{eq:gammaH2est}
\gamma_{H,s}^{(2)}\to  0 \pm
\frac{{\gamma_{H,s}^{(1)}}^2}{\gamma_{H,s}^{(0)}}\approx \pm  132.
\end{equation}
As default value for the potential coefficients in the colour-octet case, we
take the estimate obtained by the naive replacement $C_F\to (C_F-C_A/2)$ in
the known colour-singlet result~\eqref{eq:unknown-pot}. This amounts to the
numerical values
\begin{subequations}
\label{eq:cj3app}
\begin{align}
   \delta c_{J,3}^{(2,0)}|^{\text{octet}}_{\text{approx.}}&=22.2716,\\
   \delta c_{J,3}^{(1,0)}|^{\text{octet}}_{\text{approx.}}&=-8.28177
\end{align}
\end{subequations}
We estimate the uncertainty due to this estimate as $\pm 2 \delta c_{J,3}^{(n,0)}|^{\text{octet}}_{\text{approx.}}$.
It is seen  from~\eqref{eq:N3LO} that the numerical effect of the unknown
potential corrections is expected to be larger than that of the three-loop
soft-anomalous dimension.

The constants $C^{(3)}_{pp'(R)}$ in the threshold
expansion of the scaling functions~\eqref{eq:N3LO} can only be obtained
from a three-loop computation and will be set to zero in our default
approximation. The expansion of the N$^3$LL resummation formula yields
expressions for these constants that depend on the unphysical
scale parameters $k_s$, $k_h$, and $k_C$ introduced in the resummation
formalism and  are given in
Appendix~\ref{app:scaling}. 
We will use these results to estimate the uncertainty of our result
due to  effects beyond the
threshold approximation.
 Choosing as a default $k_X=1$ and varying the
scales independently in the interval $\frac{1}{2}\leq k_X \leq 2$ with
the constraint $\frac{1}{2}\leq \frac{k_s}{k_h}\leq 2$ we obtain the estimates
\begin{align}
  C^{(3)}_{q\bar q(8)}&= 
  7.0^{+25.0}_{-29.1}\times 10^3 , \label{eq:c3qq}\\
  C^{(3)}_{gg(1)}& 
  =-23.4^{+108.6}_{-78.5}\times 10^3,
 \\
  C^{(3)}_{gg(8)}&
 =-0.4^{+41.9}_{-20.8}\times 10^3 .
\end{align}
We further define the colour-averaged constant in the gluon channel,
\begin{equation}
\label{eq:c3gg}
  C^{(3)}_{gg}=\frac{\sigma^{(0),8_s}_{gg}C^{(3)}_{gg(8)}+\sigma^{(0),1}_{gg}
    C^{(3)}_{gg(1)}}{\sigma^{(0),8_s}_{gg}+\sigma^{(0),1}_{gg}}
=-6.98^{+61.0}_{-37.3}\times 10^3.
\end{equation}

\begin{figure}[t]
  \begin{center}
      \includegraphics[width=.48\textwidth]{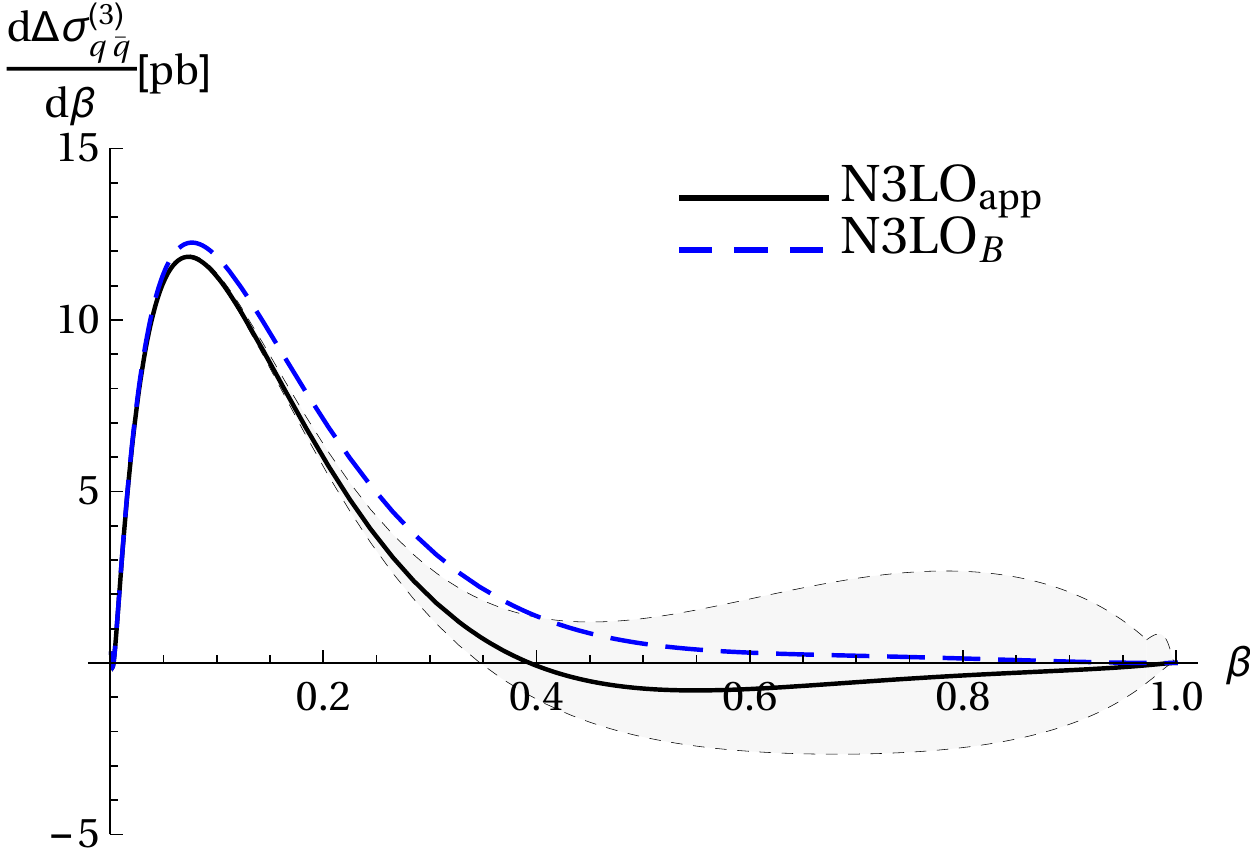}  
        \includegraphics[width=.48\textwidth]{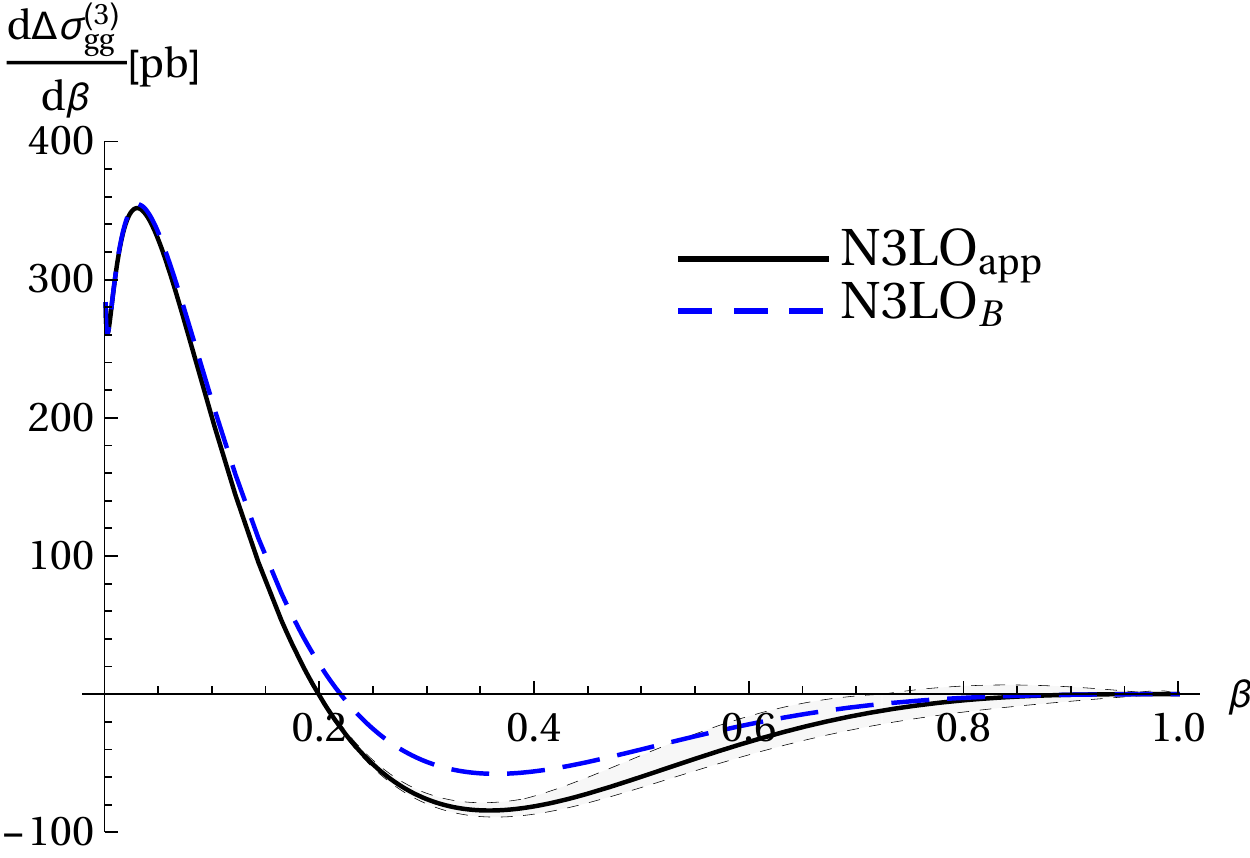}  
        \end{center}
        \caption{Partonic N$^{3}$LO corrections to the cross section for
          $q\bar q\to t\bar t$ (left) and $gg\to t\bar t$ (right) at the LHC
          ($\sqrt{s} = 13\,$TeV), multiplied with the parton
          luminosities. Black (solid): N$^3$LO$_{\text{app}}$
          using~\eqref{eq:N3LO}, where the shaded gray area indicates the
          uncertainty estimated from varying the constants $C^{(3)}_{pp'}$ and
          by using the kinematic variable $v$ instead of $\beta$.  Blue
          (long-dashed): N$^3$LO$_{\text B}$ based on NNLL resummation.}
\label{fig:N3LOpart}
\end{figure}

In Fig.~\ref{fig:N3LOpart} we compare the result~\eqref{eq:N3LO} for the
approximate N$^3$LO corrections with an earlier approximation (denoted as
N$^3$LO$_B$) based on NNLL resummation~\cite{Beneke:2011mq}, where the terms
$\sim \ln^{2,1}\beta$ and $\sim \frac{1}{\beta}$ beyond NNLL accuracy were
dropped.  In the figures we plot the integrand of the convolution of the
approximate N$^3$LO corrections~\eqref{eq:sigma3-series} with the parton
luminosity in the formula for the hadronic cross section~\eqref{eq:sig-had} as
a function of $\beta$, including the Jacobian $\partial \tau/\partial\beta$:
\begin{equation}
\label{eq:sigma-beta}
\frac{d \Delta\sigma^{(3)}_{pp'}}{d\beta}=
\frac{8\beta m_{t}^2}{s(1-\beta^2)^2}
L_{pp'}(\beta,\mu_f)\Delta\hat\sigma^{(3)}_{pp'}(\beta,\mu_f) \, .
\end{equation}
The gray band indicates the uncertainty of our default approximation~\eqref{eq:N3LO} as estimated from
the dependence of the constants~\eqref{eq:c3qq} and~\eqref{eq:c3gg} on the
unphysical scales $k_s,k_h$, and by using the kinematic variable $v$ instead
of $\beta$. In the figure the  MMHT2014NNLO
PDFs~\cite{Harland-Lang:2014zoa,Harland-Lang:2015nxa}  are used to
compute the parton luminosity.

From the comparison of the curves for our best prediction N$^3$LO$_{
\text{app}}$ to the earliear approximation N$^3$LO$_{
\text{B}}$ it can be seen that the effect of the corrections beyond NNLL is small
in the limit $\beta\to 0$ but becomes significant for
$\beta\gtrsim 0.3$. As already observed in~\cite{Beneke:2011mq}, the
corrections to the integrand $d\sigma/d\beta$ in the gluon-fusion
channel become very large for $\beta\to 0$ and could lead to
corrections of the order of several hundred picobarn to the total
cross section at the LHC with $13$~TeV.
In~\cite{Beneke:2011mq} it was speculated that this might indicate a poor convergence of the fixed-order expansion, despite an overall small numerical effect of resummation
on the total cross section.
In the total cross section,  these large positive
corrections are cancelled to a large extent by negative corrections
for $\beta >0.2$, where, however, the threshold approximation is less
reliable. The integrand approaches zero rapidly for $\beta\to 1$,
which indicates that we are at least not including spuriously large
corrections from the region where the threshold approximation breaks
down.
The gray band in the figures indicates the growing uncertainty of the
threshold approximation for $\beta\gtrsim 0.4$. The relative
uncertainty is larger in the quark-antiquark channel, however, the
impact of both channels on the uncertainty of total cross section is
similar as seen in~\eqref{eq:n3lo-is} below. 

\subsection{Phenomenological results}
\label{sec:hadronic}
In order to obtain predictions for the total hadronic top-quark pair
production cross section, we use as default the MMHT2014NNLO
PDFs~\cite{Harland-Lang:2014zoa,Harland-Lang:2015nxa} with
$\alpha_s(M_Z)=0.118$.  The top-quark pole mass value $m_t=173.3$~GeV is used.
As central factorization and renormalization scales we
choose $\mu_r=\mu_f=m_t.$
The contributions of the different partonic channels to the approximate
N$^3$LO corrections is obtained as
\begin{subequations}
\label{eq:n3lo-is}
\begin{align}
  \Delta \sigma^{\mathrm{N}^3\mathrm{LO}_{\mathrm{app}}}_{q\bar q}(13\mathrm{
    TeV}) =&1.86\mathrm{pb}\nonumber\\
  &+(0.05\, C_{q\bar q}^{(3)}- 3.91\, \delta c_{J,3}^{(1,0)}+ 3.06\, \delta c_{J,3}^{(2,0)}
 - 0.30\,\gamma_{H,s}^{(2)})\times 10^{-3}  \mathrm{pb},\\
 \Delta \sigma^{\mathrm{N}^3\mathrm{LO}_{\mathrm{app}}}_{gg}(13\mathrm{
    TeV}) =&10.09\,  \mathrm{pb} \nonumber\\
  &+(0.12\, C_{gg}^{(3)}- 7.38\, \delta c_{J,3}^{(1,0)}+ 6.15\, \delta c_{J,3}^{(2,0)}
 - 0.56\,\gamma_{H,s}^{(2)})\times 10^{-3}  \mathrm{pb},
\end{align}
\end{subequations}
where we have indicated the contributions of the various unknown coefficients
entering the scaling functions~\eqref{eq:N3LO}.  As default we set the
constant terms $C^{(3)}_{pp'}$ to zero and estimate their size according
to~\eqref{eq:c3qq} and~\eqref{eq:c3gg}. Furthermore, we estimate kinematic
ambiguities of the threshold approximation by using the expansion in terms of
the variable $v$ instead of $\beta$.  For the massive soft anomalous dimension
and the potential coefficients we choose the default values and uncertainty
estimates~\eqref{eq:gammaH2est} and~\eqref{eq:cj3app}.  We then obtain the
prediction for the N$^3$LO corrections to the cross section for
$\mu_f=\mu_r=m_t$, and the uncertainties inherent in our approximations
\begin{equation}
\label{eq:approx-error}
\Delta\sigma^{\mathrm{N}^3\mathrm{LO}_{\mathrm{app}}}_{t\bar t}(13\mathrm{
    TeV})= 12.25
  \underbrace{{}_{-6.24}^{+7.87}}_{C_{pp'}^{(3)}}
  \underbrace{{}^{+5.3}_{-0.0}}_{\mathrm{kin.}}  
  \pm \underbrace{0.11}_{\gamma_{H,s}^{(2)}}
 \pm\underbrace{0.60}_{\delta c_{J,3}^{(i,0)}}\mathrm{pb},
\end{equation}
where ``kin.'' denotes the uncertainty due to the expansion in terms of $v$
instead of $\beta$. It is clearly seen that the uncertainties of the threshold
expansion as estimated by the variation of the constants and the kinematic
ambiguity dominate over the uncertainties due to the unknown soft anomalous
dimension and potential coefficients. 

In addition to the corrections included in the scaling
functions~\eqref{eq:N3LO}, there is the distributional contribution arising
from the third Coulomb correction~\eqref{eq:c3dist}. Inserting this correction
to the potential function into the factorization formula~\eqref{eq:fact} gives
the correction\footnote{The present notation is related to the one
  in~\cite{Beneke:2016jpx} by $ H^{R}_{pp',i}=\frac{2\pi^2\alpha_s^2}{m_t^4}\sigma_{pp'}^R$.}
\begin{equation}
\label{eq:sigma-c3}
  \Delta\sigma_{N_1N_2\to t\bar tX}^{C_3}=
  \sum_{p,p'=q,\bar q,g}\,L_{pp'}( \frac{4m_t^2}{s},\mu_f) \sum_{R=1,8}
\frac{ \alpha_s^3(\mu_r)(- D_R)^3m_t^4\zeta_3 }{2s}  H^{R}_{pp',i}(m_t,\mu_r).
\end{equation}
The resulting correction at the LHC with $\sqrt{s}=13$~TeV is given by
\begin{equation}
    \Delta\sigma_{t\bar t}^{C_3}(13\mathrm{TeV})=0.60\,\mathrm{pb},
\end{equation}
in agreement with~\cite{Beneke:2016jpx}. This contribution to the N$^3$LO
cross section is therefore of a similar size as the uncertainties due to the
unknown coefficients in  N$^3$LL resummation.

Our default prediction is obtained by adding the correction
$\Delta\sigma^{\mathrm{N}^3\mathrm{LO}_{\mathrm{app}}}_{t\bar t}$ from the
threshold expansion of the scaling functions and the
correction~\eqref{eq:sigma-c3} to the NNLO cross section~\cite{Baernreuther:2012ws,Czakon:2012zr,Czakon:2012pz,Czakon:2013goa},
\begin{equation}
   \sigma^{\mathrm{N}^3\mathrm{LO}_{\mathrm{app}}}_{t\bar t}=
   \sigma^{\mathrm{NNLO}}_{t\bar
     t}+\Delta\sigma^{\mathrm{N}^3\mathrm{LO}_{\mathrm{app}}}_{t\bar t}
   + \Delta\sigma_{t\bar t}^{C_3}.
\end{equation}
In order to estimate the scale uncertainty, we vary both renormalization
and factorization scales independently in the interval $\frac{1}{2}\leq
\mu_f,\mu_r  \leq 2$ with the constraint $\frac{1}{2}\leq
\frac{\mu_f}{\mu_r}\leq 2$.
We obtain
\begin{equation}
\label{eq:n3lo-13}
  \sigma^{\mathrm{N}^3\mathrm{LO}_{\mathrm{app}}}_{t\bar t}(13\mathrm{
    TeV})=815.70^{+19.88 (2.4 \%)}_{-27.69 (3.4 \%)} (\mathrm{scale}) 
    {}^{+9.49 (1.2 \%)}_{-6.27 (0.8 \%)}(\mathrm{approx})
\mathrm{pb},
\end{equation}
where the ``approx'' uncertainty is obtained by adding the different
contributions in~\eqref{eq:approx-error} in quadrature.
Compared to the NNLO result,
\begin{equation}
  \sigma^{\mathrm{NNLO}}_{t\bar t}(13\mathrm{
    TeV})=802.85^{+28.12 (3.5\%)}_{-44.97 (5.6 \%)} (\mathrm{scale})\mathrm{pb},
\end{equation}
the approximate N$^3$LO corrections increase the cross section moderately by
$1.6\%$ and reduce the scale uncertainty to the $\pm 3\% $-level.  Adding
scale and approximation uncertainty in quadrature leads to a total
perturbative uncertainty of ${}^{+2.7 \%}_{-3.4 \%}$, and therefore a
reduction compared to the scale uncertainty of the NNLO result. 

For other centre-of-mass energies at the LHC we obtain the results
\begin{align}
\sigma^{\mathrm{N}^3\mathrm{LO}_{\mathrm{app}}}_{t\bar t}(7\mathrm{
    TeV})&=  175.56^{+4.45 (2.5\%)}_{-5.85 (3.3\%)} (\mathrm{scale})
  {}^{+2.15 (1.2 \%)}_{-1.45 (0.8 \%)}(\mathrm{approx})\mathrm{pb}, \label{eq:n3lo-7}\\
\sigma^{\mathrm{N}^3\mathrm{LO}_{\mathrm{app}}}_{t\bar t}(8\mathrm{
    TeV})&=  250.22_{-8.38(3.5\%)}^{+6.30(2.5\%)} (\mathrm{scale})
  {}^{+3.03 (1.2 \%)}_{-2.03(0.8 \%)}(\mathrm{approx})\mathrm{pb}, \\
\sigma^{\mathrm{N}^3\mathrm{LO}_{\mathrm{app}}}_{t\bar t}(14\mathrm{
    TeV})&=  964.32_{-32.8(3.4\%)}^{+23.4(2.4\%)} (\mathrm{scale})
  {}^{+11.15(1.2 \%)}_{-7.36(0.8 \%)}(\mathrm{approx})\mathrm{pb} .
 \label{eq:n3lo-14}
\end{align}

 In addition
to the uncertainty of the perturbative calculation estimated by the scale
variation, the prediction of the hadronic cross section also relies on the
value of the strong coupling constant and the PDFs, with their respective
uncertainties. For the MMHT2014NNLO PDF, the combined PDF+$\alpha_s$
uncertainty of the NNLO top-quark pair-production cross section at the LHC is
about $\pm 2.5\%$~\cite{Harland-Lang:2015nxa}, and therefore below the
perturbative uncertainty of the approximate N$^3$LO result~\eqref{eq:n3lo-13}.
Following~\cite{Anastasiou:2016cez} we note that the evolution of the NNLO
PDFs is performed including the three-loop splitting function, so the use of
NNLO PDFs in an N$^3$LO calculation is formally consistent, although the use
of N$^3$LO predictions in a PDF fit might lead to a non-negligible  effect on
the PDFs, in particular in the case of top-pair
production~\cite{Forte:2013mda}. In the following we discuss only the estimate
of the  perturbative uncertainties of our calculation.

It is  interesting to compare the approximate N$^3$LO result to
the prediction of NNLL resummation matched to NNLO. Using the programs
\texttt{topixs 2.0.1}~\cite{Beneke:2012wb} and
\texttt{top++ 2.0}~\cite{Czakon:2011xx}, which implement the combined 
soft-gluon and Coulomb resummation in momentum
space~\cite{Beneke:2011mq} and soft-gluon resummation in Mellin
space~\cite{Cacciari:2011hy}, respectively, we obtain
\begin{align}
    \sigma^{\mathrm{NNLL+NNLO}}_{t\bar t}(13\mathrm{
    TeV})&=807.13_{-36.83 (4.6\%)}^{+15.63 (1.9\%)} (\mathrm{scale})
           {}_{-12.9 (1.8\%)}^{+19.15 (2.5\%)} (\mathrm{approx.})\mathrm{pb} &
    \mathrm{\texttt{(topixs)}},\\
    \sigma^{\mathrm{NNLL+NNLO}}_{t\bar t}(13\mathrm{
    TeV})&=  821.37^{+ 20.28 (2.5\%)}_{- 29.60(3.6\%)}
           (\mathrm{scale}) \mathrm{pb}
&\mathrm{\texttt{(top++)}}.
\end{align}
 The resummed Coulomb
corrections beyond NNLO and bound-state corrections included in
\texttt{topixs} only have a small effect, $+2.8$~pb for the bound-state
corrections and less than one picobarn for higher-order Coulomb corrections.
Furthermore, \texttt{top++} includes the two-loop
constants $C_{pp'}^{(2)}$, which are part of the NNLL' approximation. Setting
these constants to zero, one obtains the \texttt{top++} result of $812.20$~pb,
which is closer to the result of \texttt{topixs}. The remaining difference
indicates the size of sub-leading corrections, which are treated differently
in the different resummation methods. 
The \texttt{top++} result is also
closer to the approximate N$^3$LO prediction~\eqref{eq:n3lo-13}, which
indicates the numerical relevance of the interplay of the two-loop constants
and the soft threshold corrections.  The total scale uncertainty of the two
NNLL results is similar to each other and to that of the approximate N$^3$LO
correction.  The \texttt{topixs} result also includes an estimate of the
ambiguities of the resummation procedure, which includes the variation of the
soft scale used in the momentum-space approach and the comparison of the
expansion in $v$ and $\beta$.  Since our approximate N$^3$LO cross section is
based on a partial N$^3$LL resummation, the dependence on the soft-scale is
reduced compared to the NNLL resummation and enters only in the estimate of the
constant term.  No attempt to estimate the ambiguities of the resummation
procedure is made in \texttt{top++}.
The fact that the approximate N$^3$LO corrections are close to the resummed
results indicates that the perturbative expansion is well-behaved, despite the
large corrections to the partonic cross sections for small $\beta$ observed in
Figure~\ref{fig:N3LOpart}.

\begin{figure}[t]
  \begin{center}
      \includegraphics[width=.6\textwidth]{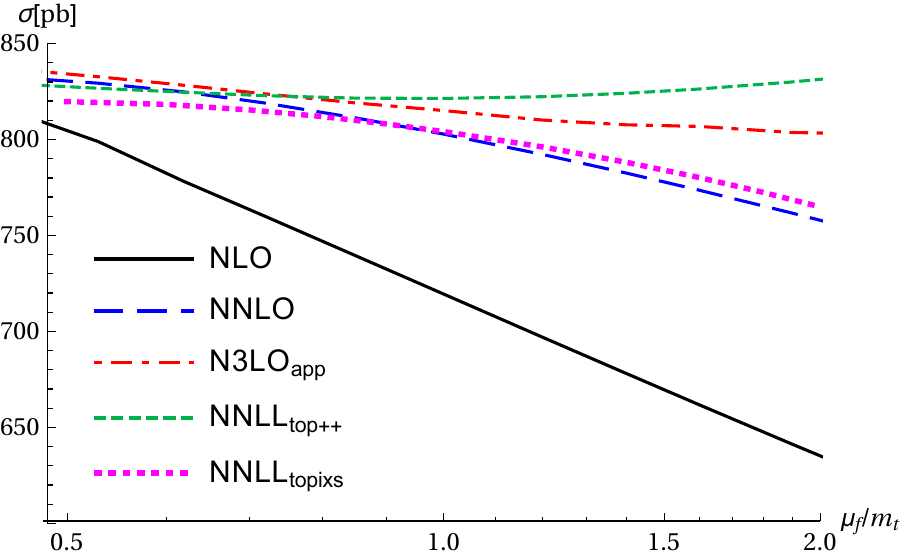}  
         \end{center}
\caption{Scale-dependence ($\mu_f=\mu_r$) of the total cross section for the
  LHC with $\sqrt{s}=13$~TeV in various approximations.}
\label{fig:N3LOmu}
\end{figure}
\begin{figure}[t]
  \begin{center}
    \includegraphics[width=.49\textwidth]{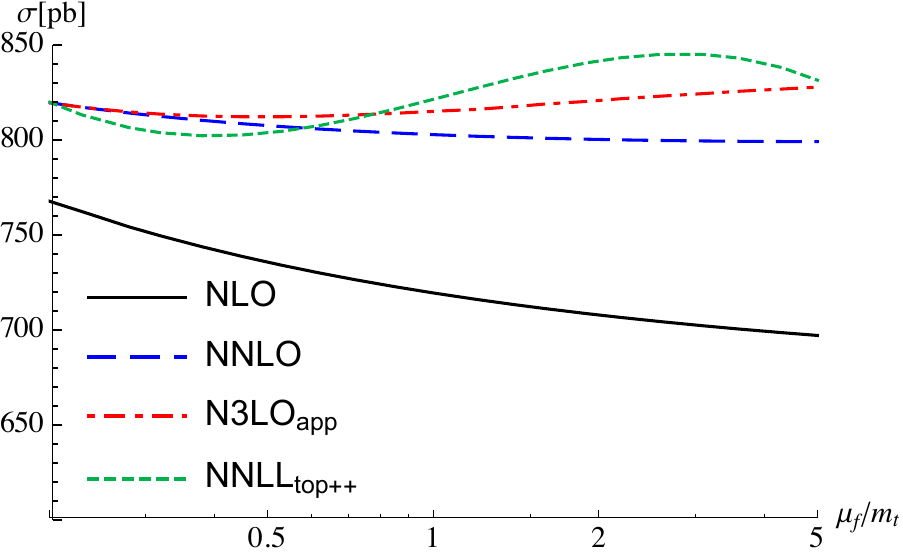}  
      \includegraphics[width=.49\textwidth]{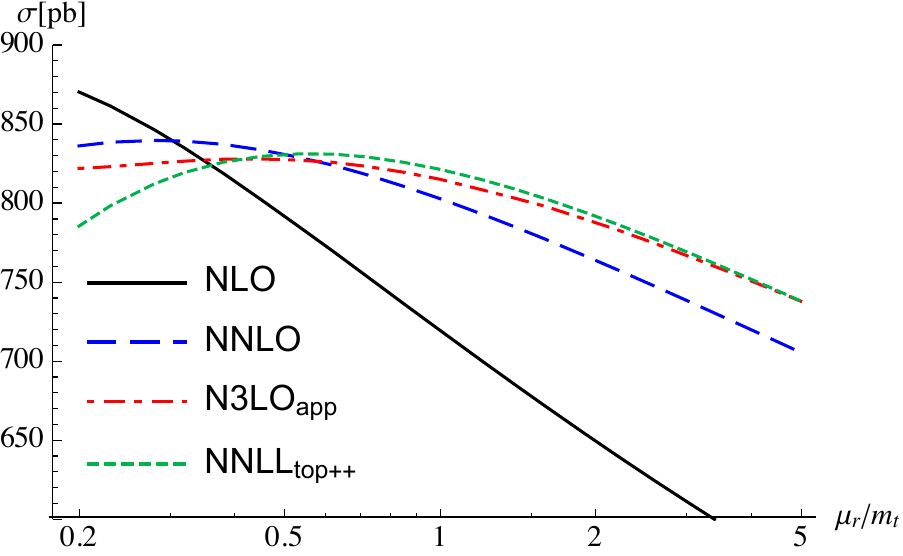}  
         \end{center}
\caption{Total cross section for the
  LHC with $\sqrt{s}=13$~TeV in various approximations. Left:
  factorization-scale dependence (for $\mu_r=m_t$), right: renormalization-scale dependence (for $\mu_f=m_t$).}
\label{fig:N3LOmufr}
\end{figure}

In Figure~\ref{fig:N3LOmu} we compare the factorization-scale dependence of
the approximate N$^3$LO cross section (red dash-dotted) to the NNLO result
(blue, long-dashed) and the two resummed predictions from \texttt{top++}
(green, dashed) and \texttt{topixs} (magenta, dotted). The renormalization
scale is set equal to the factorization scale. In the \texttt{topixs}
calculation, all other scales are fixed to their default value. It is seen
that the inclusion of the approximate N$^3$LO corrections further flatten the
scale dependence compared to the NNLO prediction and brings the curve closer to
the resummed \texttt{top++} result. In contrast, the  \texttt{topixs} curve
follows more closely the NNLO result.
 In Figure~\ref{fig:N3LOmufr} we consider instead the
factorization scale dependence for fixed renormalization scale and
vice-versa. No results from \texttt{topixs} are shown since the
renormalization scale is not separated from the factorization scale. Again it
is seen that the inclusion of the approximate N$^3$LO corrections stabilize
the scale dependence and brings the result closer to the resummed
prediction.

Approximate N$^3$LO results were also published
in~\cite{Kidonakis:2014isa,Muselli:2015kba}. In these papers, no analytical
results for the partonic cross sections are given so we can only compare
numerical results for the total cross section. In addition, different PDFs are
used for the results.  The result of~\cite{Kidonakis:2014isa} is based on NNLL
resummation for one-particle inclusive kinematics. At the LHC with $13$~TeV, a
correction of $2.7\%$ relative to the NNLO result is found with a scale
uncertainty of ${}^{+2.9\%}_{-2.0\%}$. The inherent uncertainty of the
constructed approximation is not estimated.  Ref.~\cite{Muselli:2015kba}
constructs a partial N$^3$LL soft resummation in Mellin space using similar
input as our approximation, but without the contributions of the N$^3$LO
potential function. Furthermore, they also include $1/N$-suppressed
contributions in Mellin space and information on the large-$\beta$
behaviour. As a result, they find larger corrections of $+4.2\%$ at
$13$~TeV. The scale uncertainty of $\pm 2.7\%$ is comparable to that of our
result~\eqref{eq:n3lo-13}.  The scale-dependence of our result in
Figure~\ref{fig:N3LOmufr} is qualitatively similar to that shown in Figure 9
of~\cite{Muselli:2015kba}.  An approximation corresponding to the standard
Mellin-space resummation~\cite{Cacciari:2011hy,Czakon:2011xx} dubbed
``$N$-soft'' yields smaller corrections of $+2.3\%$ and is therefore closer to
our prediction.  The ambiguities of the approximation are estimated as $\pm
1.9\%$, which is larger than our estimate of ${}^{+1.2\%}_{-0.8\%}$
in~\eqref{eq:n3lo-13}.  Therefore, while our default implementation results in
a smaller correction than the one found
in~\cite{Kidonakis:2014isa,Muselli:2015kba}, the results are consistent if the
estimate of the uncertainties due to the threshold approximation are taken
into account. In particular, using the expansion in terms of $v$ instead of
$\beta$, we obtain a larger N$^3$LO effect of $2.3\%$ which is in good
agreement with the ``$N$-soft'' prediction of~\cite{Muselli:2015kba} and
with~\cite{Kidonakis:2014isa}.  
It is interesting to
compare our uncertainty estimate in more detail to the one
of~\cite{Muselli:2015kba}.  The ``approx'' uncertainty estimate
in~\cite{Muselli:2015kba} includes contributions from two sources: the
treatment of the subleading contributions in Mellin space and the parametric
uncertainties due to the unknown three-loop constant term and the massive soft
anomalous dimension for the colour-octet case. The estimate of the three-loop
constant in Eqs.~(5.3) and~(5.5) of~\cite{Muselli:2015kba} amounts to an
uncertainty of the cross section of $\pm 2.6$ pb at the LHC, which is smaller
than our estimate in~\eqref{eq:approx-error}. In contrast, the estimate of the
colour-octet soft-anomalous dimension in~\cite{Muselli:2015kba} amounts to an
uncertainty of $\pm 4.6$ pb which is much more conservative than our
estimate.\footnote{The conventions of~\cite{Muselli:2015kba} are related to
  our conventions according to $(4\pi)^3(D_{\bf 8}^{(3)}-D_{\bf
    1}^{(3)})=2C_A\gamma_{H,s}^{(2)}+\dots$, where
  additional terms arise due to the two-loop soft
  function~\cite{Beneke:2009rj}. The uncertainty estimate of $(D_{\bf
    8}^{(3)}-D_{\bf 1}^{(3)})=\pm 10$ is therefore much larger
  than~\eqref{eq:gammaH2est}. } Nevertheless, combining the two sources of
parametric uncertainty in quadrature results in a similar parametric
uncertainty as in our result. The dominant contribution to the uncertainty
estimate of~\cite{Muselli:2015kba} arises from the ambiguities due to the
treatment of the subleading contributions in Mellin space, which contribute an
uncertainty of $\pm 1.8\%$, compared to our smaller estimate of the kinematic
ambiguities using the expansion in $v$ instead of $\beta$.  In addition,
Ref.~\cite{Muselli:2015kba} estimates the uncertainty from omitting the
N$^3$LO approximation for the quark-antiquark and quark-gluon initial states
as $1\%$. Instead, our result~\eqref{eq:n3lo-is} includes the quark-antiquark
channel with a contribution of only $0.2\%$.
 
\section{Conclusions}
We have constructed a combined resummation of threshold logarithms and Coulomb
corrections to the total top-quark pair-production cross section at hadron
colliders at partial N$^3$LL accuracy and used the result to compute the
threshold limit $\beta=\sqrt{1-\frac{4m_t^2}{\hat s}}\to 0$ of the partonic
N$^3$LO cross sections $q\bar q \to t\bar t$ and $gg\to t\bar t$. Our result
makes use of the state-of-the-art N$^3$LO results for the non-relativistic
Green function from a calculation of $e^-e^+\to t\bar t$. We have also
included contributions from sub-leading P-wave production channels and
power-suppressed so-called next-to-eikonal corrections that give rise to
threshold-enhanced N$^3$LO corrections due to Coulomb enhancement.  We have
generalized the so-called $1/m^2$ one-loop potential in PNRQCD to the
spin-singlet and colour-octet state.  However, the three-loop massive
soft-anomalous dimension, the two-loop $1/m$ potential in PNRQCD and the
ultrasoft corrections to the non-relativistic Green function are currently not
known for the production of top-quark pairs in a colour-octet state. We have
estimated the uncertainty due to these missing ingredients on our result.  The
corrections from the N$^3$LO Green function and the missing contributions for
the colour-octet state affect the coefficients of the terms
$\alpha_s^3\ln^{2,1}\beta$, and can have a larger numerical effect than the
unknown three-loop massive soft-anomalous dimension.

Using the results for the threshold expansion of the partonic N$^3$LO cross
sections, we obtain moderate corrections $+1.5\%$ relative to the NNLO
predictions for the LHC with $\sqrt{s}=13$~TeV and a reduction of the
factorization- and renormalization-scale uncertainty to the $\pm 3\%$-level.
We estimate the uncertainty of our prediction due to the threshold
approximation to be about $\pm 1\%$. The total perturbative  uncertainty 
therefore approaches a similar level as the PDF+$\alpha_s$ uncertainty of $\pm
2.5\%$. Our default prediction is smaller than other approximate N$^3$LO
predictions~\cite{Kidonakis:2014isa,Muselli:2015kba} but consistent with these
results if the estimate of the uncertainties due to the threshold
approximation are taken into account. The difference between these predictions
indicates the possible size of subleading terms in the threshold expansion.

 We have explicitly provided all necessary ingredients to implement
our results in a numerical program. 
Since the numerical result of the approximate N$^3$LO corrections and
their scale uncertainty are close to the resummed NNLL predictions, this
provides a simple, computationally less expensive way to obtain the
leading corrections beyond NNLO. We plan to include these results in a
future version of \texttt{topixs}, as well the implementation of the fully
resummed threshold corrections at partial N$^3$LL accuracy.

\subsection*{Acknowledgments}
We would like to thank Martin Beneke for useful discussions and for making a
copy of Ref.~\cite{Wuester:2003} available.
The work of CS is supported by the Heisenberg Programme of the DFG.
We acknowledge support by the Munich Institute for Astro- and Particle
Physics (MIAPP) of the DFG cluster of excellence ``Origin and Structure
of the Universe''  during parts of this work.
\subsection*{Note added}
In this version v3 of the arXiv submission of the paper, typos in the
coefficients of the $\ln\beta$-terms in the scaling functions
$f^{(3,0)}_{gg(1)}$ in~\eqref{eq:N3LO}, $f^{(3,1)}_{gg(8)}$
in~\eqref{eq:f31gg8}, and the $1/\beta$-term in $f^{(3,2)}_{gg(1)}$
in~\eqref{eq:f32gg1} have been corrected. These typos do not affect the
implementation used to compute the
numerical results of the cross sections. However, in addition the scale dependence in the
results~\eqref{eq:n3lo-13} and~\eqref{eq:n3lo-7}--\eqref{eq:n3lo-14} has been
corrected.

In the latest version \texttt{3.0} of the program
\texttt{topixs}, which can be obtained from
\begin{center}
\url{http://users.ph.tum.de/t31software/topixs/}
\end{center}
the approximate N$^3$LO corrections computed in this paper are available
as an option  by setting \texttt{APPROX\_NNNLO=1} in the input file \texttt{topixs.cfg}.
\appendix
\section{Renormalization group functions}
\label{app:input}
In this appendix we give explicit results for the NNLO expansion of the hard and soft functions.
The various
anomalous dimensions appearing in the expressions are expanded in the strong coupling constant according to
\begin{equation}\label{eq:gamma-alpha}
\gamma=\sum_n \gamma^{(n)}
\left(\frac{\alpha}{4\pi}\right)^{n+1}.
\end{equation}
 For the expansion of the $\beta$-function and
the splitting functions we use the conventions
\begin{align}
\label{eq:beta-alpha}
  \beta(\alpha_s)&=-\alpha_s\sum_{n=0}^\infty \beta_{n}
  \left(\frac{\alpha_s}{4\pi}\right)^{n+1}\\
  P_{p/p'}(x)&=\sum_{n=0}^\infty P_{p/p'}^{(n)}(x)
  \left(\frac{\alpha_s}{4\pi}\right)^{n+1}
\label{eq:p-alpha}
\end{align}
which corresponds to $\beta_0= \frac{11}{3} C_A-\frac{4}{3} n_l T_f$.

\subsection{Two-loop hard functions}
\label{app:h2}

 The perturbative expansion of the hard function in 
the QCD coupling can be written as
\begin{equation}
\label{eq:hard-def}
   H^{R}_{pp'}(\mu_h)= H^{R(0)}_{pp'}(\mu_h)
   \left[1+\sum_n\left(\frac{\alpha_s(\mu_h)}{4\pi}\right)^n h^{R(n)}_{pp'}(\mu_h)
     \right] \, ,
\end{equation}
where we have extracted the leading-order hard function
$H^{R(0)}_{pp'}(\mu_h)$ that includes a factor $\alpha_s^2(\mu_h)$. We also
left the spin dependence implicit.  The relation of the one- and two-loop
coefficients at a scale $\mu$ to the initial conditions at $\mu_h=m_t$ follow
from the evolution equation~\eqref{eq:rge-hard}:
\begin{align}
h^{R(1)}_{pp'}(\mu)&=-C_{rr'}\gamma^{(0)}_{\text{cusp}}
\ln^2\left(\frac{m_t}{\mu}\right)    
-2(\gamma^{V(0)}_i+2\beta_0 +C_{rr'}\gamma^{(0)}_{\text{cusp}}\ln 2)
\ln\left(\frac{m_t}{\mu}\right)
+h^{R(1)}_{pp'}(m_t),\\
h^{R(2)}_{pp'}(\mu)&= \frac{1}{2}C_{rr'}^2(\gamma^{(0)}_{\text{cusp}})^2
\ln^4\left(\frac{m_t}{\mu}\right)
+2C_{rr'}\gamma^{(0)}_{\text{cusp}}\left(\gamma^{V(0)}_i+C_{rr'}\gamma^{(0)}_{\text{cusp}}
  \ln 2+\frac{7}{3}\beta_0\right)
\ln^3\left(\frac{m_t}{\mu}\right)    \nonumber\\
&\quad
+\Bigl[\left(\gamma^{V(0)}_i+2\beta_0\right)
  \left(2\gamma^{V(0)}_i+6\beta_0 + 5 C_{rr'}\gamma^{(0)}_{\text{cusp}}\ln
    2\right)
 \nonumber\\
&\quad
  -C_{rr'}
  \left(\gamma^{(1)}_{\text{cusp}}+\gamma^{(0)}_{\text{cusp}}h^{R(1)}_{pp'}(m_t)
  -2C_{rr'}(\gamma^{(0)}_{\text{cusp}})^2\ln^2 2+ \gamma^{V(0)}_i\gamma^{(0)}_{\text{cusp}}\ln 2\right)
\Bigr]
\ln^2\left(\frac{m_t}{\mu}\right)\nonumber\\
&\quad -\Bigl[2\gamma^{V(1)}_i+\gamma_J^{R,S(1)} +2C_{rr'}\gamma^{(1)}_{\text{cusp}}\ln 2
  +4\beta_1\nonumber\\
&\quad +2\left(\gamma^{V(0)}_i+3\beta_0 +C_{rr'}\gamma^{(0)}_{\text{cusp}}\ln 2\right)h^{R(1)}_{pp'}(m_t)
\Bigr]\ln\left(\frac{m_t}{\mu}\right)+h^{R(2)}_{pp'}(m_t).
\end{align}
Here the factor $\alpha_s^2(\mu_h)$ in the leading-order hard function was
taken into account.  We have also defined the colour factor $C_{rr'}=C_r+C_{r'}$.

Expanding the resummation formula for the NNLL and NNLL' approximations to
$\mathcal{O}(\alpha_s)$ and $\mathcal{O}(\alpha_s^2)$, respectively, one
obtains the relation between the constant terms $C_{pp}^{R(n)}$ in the
threshold expansion of the NLO and NNLO cross sections to the coefficients of
the hard functions at the scale $\mu_h=m_t$:
\begin{align}
\label{eq:c1}
  h^{R(1)}_{pp'}(m_t)&=C_{pp}^{R(1)}
  -C_{rr'}\left(32+36\ln^22-48\ln 2-\frac{11\pi^2}{6} \right)
  -12C_R(1-\ln 2), \\
  h^{R(2)}_{pp'}(m_t)&=C_{pp}^{R(2)}-\tilde s_i^{R(2)}(0) 
  -(4\pi)^2  (c_{J,2} +\gamma_J^{R,S(1)}\ln 2)
-  \frac{4\pi^4  D_R^2}{3}\left[ d_J-\frac{3}{8}\right] \nonumber\\
 &\quad  
  -h^{R(1)}_{pp'}(m_t)\Biggl[C_{rr'}\left(4(8+9\ln^22)
      -48\ln 2-\frac{11 \pi ^2}{6}\right) 
    +12 C_R(1-\ln 2)\Biggr]
  \nonumber\\
&\quad
  -C_{rr'}^2\Biggl[24
   \left(27 \ln^4 2-72 \ln^3 2+144 \ln^2 2-192
   \ln 2+128\right)\nonumber\\
&\quad   
    -\frac{70 \pi ^2}{3}\left(9 \ln^2 2-12 \ln 2+8\right)
    -\frac{7\pi^4}{3}+448(3\ln 2-2)\zeta_3\Biggr]
  \nonumber\\
&\quad -C_{rr'}C_R\left[-16 \left(27 \ln^3 2-63 \ln^2 2+84
   \ln 2-56\right)-\frac{\pi ^2}{3} (164-210 \ln 2)-224
   \zeta_3 \right]
  \nonumber\\
&\quad  -C_{rr'}C_A \Bigl[268 \ln^2 2-\frac{4024
  \ln 2}{9}+\frac{8048}{27}+\frac{2 \pi ^4}{3}\nonumber\\
&\quad 
-\frac{1}{3} \pi ^2 \left(36 \ln^2 2-59 \ln 2+84\right)+28 (3 \ln 2-2) \zeta_3\Bigr]
  \nonumber\\
 &\quad -C_{rr'} n_f T_F \left[
\frac{8}{27}\left(-270\ln^22+444 \ln 2 -296\right)
+\frac{ 4\pi^2}{9}(12-3\ln 2)\right] \nonumber\\
 &\quad 
 -C_R^2\left[24\left(3\ln^22-6\ln 2+4\right)-4 \pi ^2\right]  
\nonumber\\
 &\quad 
 -C_R\Bigl[C_A\left(\frac{196}{9}
   -\frac{4}{3}\pi ^2
   +8 \zeta_3\right)(2-3\ln 2)
 -n_f T_F\frac{1}{9}\left(160-240\ln 2\right)\Bigr]
 \nonumber\\
 &\quad
 -\beta_0  \Biggl[C_{rr'} \left( 
-24 (3\ln^2 2-6\ln 2+8) \ln 2 +11 \pi ^2 \ln 2
-\frac{112 \zeta_3}{3}-\frac{22 \pi ^2}{3}+128
\right)\nonumber\\
 &\quad
+12 C_R\left(3 \ln^22-6 \ln 2+4 -\frac{\pi^2}{6}\right) \Biggr].
\label{eq:c2}
\end{align}
Although the natural hard scale is of the order $2m_t$, these relations
simplify for the choice $\mu_h=m_t$ made here. In particular, they become
independent of the anomalous dimensions $\gamma^V_i$, which do not obey
Casimir scaling. In the NNLO coefficient, the constant terms in the two-loop
soft function~\eqref{eq:s20} and the NNLO potential function~\eqref{eq:cj2}
enter.  Additionally, the coefficient $d_J$ from kinematic corrections to the
matching of the potential function~\eqref{JRal0} and the
relation~\eqref{eq:vbeta} of the variables $v$ and $\beta$ have been taken
into account.  The constants $C_{pp'}^{R(2)}$ in~\eqref{eq:c2} are related to
the results in the conventions of~\cite{Baernreuther:2013caa}, here denoted by
a bar, according to
\begin{equation}
\label{eq:cbar}
 C_{pp'}^{R(2)}= \bar C_{pp'}^{R(2)}+\frac{4\pi^4 D_R^2}{3}
- 4\pi^2 D_R^2\left(1+\frac{\pi^2}{3}\right) h^{(0),R}_{pp',P}.
\end{equation}
Here the first correction term arises from multiplying the prefactor
$4m_t^2/\hat s=1-\beta^2$ with the second Coulomb correction.  Further, since
no spin decomposition is performed in the results for the constants $\bar
C_{pp}^{R(2)}$ in~\cite{Baernreuther:2013caa}, the P-wave
contribution~\eqref{eq:pwave-nnlo} with the coefficient~\eqref{eq:hP} was
subtracted to obtain the hard function for the S-wave process.

For the number of
light flavours $n_f=5$ and using $N_C=3$ and $T_F=1/2$, but keeping the colour
representations arbitrary, the NNLO result becomes
\begin{align}
   h^{R(2)}_{pp'}(m_t)&=\bar C_{pp'}^{R(2)}-\tilde s_i^{R(2)}(0) 
 +h^{R(1)}_{pp'}(m_t)(2.06903 C_{rr'}-3.68223 C_R)   \nonumber\\
  &\quad
-5.41462 C_{rr'}^2  + 25.0828 C_{rr'} C_R + 
 35.5195 C_{rr'} + 8.69899 C_R^2 + 35.9266 C_R\nonumber\\
 &\quad
+D_R^2 (-554.062+61.0012 \,\nus-78.9568\, \nuseps-129.879\, d_J-169.357
h^{(0),R}_{pp',P}  ) \nonumber\\
  &\quad
+D_R(263.804 +15.2503 {\nua}^{(0)}) .
\label{eq:h2-exp}
\end{align}

\subsection{Two-loop soft function}
\label{app:soft}

We consider the perturbative expansion of the soft function in Laplace space,
\begin{equation}
\label{eq:soft-alpha}
   \tilde s_i^{R}(\rho,\mu)=\sum_{n=0}^\infty
 \left(\frac{\alpha_s(\mu)}{4\pi}\right)^{n}  
\tilde s_i^{(n)R}(\rho,\mu).
\end{equation}
The coefficients of the variable $\rho$ are fixed by the
anomalous dimensions in the evolution equation~\eqref{eq:rge-soft}, so the
only required input are constant coefficients $\tilde s_i^{(n)R}(0)$:
\begin{align}
 \tilde s_i^{(0)R}(\rho,\mu)&= 1, \\
 \tilde s_i^{(1)R}(\rho,\mu)&= 
 \frac{C_{rr'}\gamma_{\text{cusp}}^{(0)}}{4}\, \rho^2 + C_R\gamma_{H,s}^{(0)}\, \rho
+ \tilde s_i^{(1)R}(0),
\label{eq:soft-one}\\
 \tilde s_i^{(2)R}(\rho,\mu)&= 
 \frac{1}{2}\left(\frac{C_{rr'}\gamma_{\text{cusp}}^{(0)}}{4}\right)^2\, \rho^4
+\frac{C_{rr'}\gamma_{\text{cusp}}^{(0)}}{12}(3 C_R\gamma_{H,s}^{(0)}-\beta_0)\, \rho^3\nonumber\\
&\quad + \frac{1}{4}\left[C_{rr'}(
  \gamma_{\text{cusp}}^{(1)}+\gamma_{\text{cusp}}^{(0)} \tilde s_i^{(1)R}(0))
+2C_R\gamma_{H,s}^{(0)}( C_R\gamma_{H,s}^{(0)}-\beta_0)\right]\rho^2\nonumber\\
&\quad +\left[ C_{rr'} \gamma_s^{(1)}+C_R\gamma_{H,s}^{(1)}
  +\tilde s_i^{(1)R}(0)(C_R\gamma_{H,s}^{(0)}-\beta_0) \right]\rho   + 
\tilde s_i^{(2)R}(0).
\end{align}
Here it was used that the soft anomalous dimensions of the light partons
$\gamma_s$ vanish at the one-loop level.  The initial conditions $\tilde
s_i^{(2)R}(0)$ given in~\eqref{eq:s20} were obtained from two-loop
calculations of the soft function in the literature by noting that the
soft function in Laplace space is obtained from the position-space result
in~\cite{Becher:2007ty} by the simple replacement $L=2\ln\left(\frac{i z_0\mu
    e^{\gamma_E}}{2}\right)\to -\rho$ and from the Mellin-space result
in~\cite{Czakon:2013hxa} by the replacement $L=\ln\left(\frac{\mu
    N}{m_t}\right)\to -2 \rho$.

The explicit results for the cusp anomalous dimension are
given by
\begin{align}
\gamma^{(0)}_{\text{cusp}}&= 4 , \nonumber\\
\gamma^{(1)}_{\text{cusp}}&= 
\left(\frac{268}{9}-\frac{4\pi^2}{3}\right)C_A
-\frac{80}{9} T_F n_f ,  \nonumber \\
\gamma^{(2)}_{\text{cusp}}&=C_A^2\left(\frac{490}{3}-\frac{536\pi^2}{27}+\frac{44\pi^4}{45}+\frac{88}{3}\zeta_3\right)
+C_A T_F
n_f\left(-\frac{1672}{27}+\frac{160\pi^2}{27}-\frac{224}{3}\zeta_3\right)\nonumber\\
&\qquad+C_FT_Fn_f\left(-\frac{220}{3}+64\zeta_3\right)-\frac{64}{27}T_F^2
n_f^2.
\label{eq:cusp}
\end{align}

The one- and two-loop coefficients of the soft anomalous dimensions of the
light partons are given in~\cite{Becher:2007ty} while the three-loop
coefficient can be obtained from the results for $\gamma^{(2)p}$ and
$\gamma^{(2)\phi,p}$ in ~\cite{Becher:2009qa,Becher:2007ty,Ahrens:2008nc}
\begin{align}
 \gamma_s^{(0)}&=0 , \nonumber\\
  \gamma_s^{(1)}&=C_A
  \left(-\frac{404}{27}+\frac{11\pi^2}{18}+14\zeta_3\right)
  +T_F n_f\left(\frac{112}{27}-\frac{2\pi^2}{9}\right), \nonumber\\
  \gamma_s^{(2)}&=C_A^2\left(-\frac{136781}{1458}+\frac{6325 \pi ^2}{486}
-\frac{44 \pi^4}{45}+\frac{658}{3}\zeta_3-\frac{44 \pi^2}{9} \zeta_3
-96 \zeta_5\right)\nonumber\\
&+ C_A T_F n_f\left(\frac{11842}{729}-\frac{1414 \pi
   ^2}{243}+\frac{8 \pi ^4}{15}-\frac{728}{27} \zeta_3\right)
\nonumber\\
&+C_F T_F n_f\left(\frac{1711}{27}-\frac{2 \pi ^2}{3}-\frac{8
   \pi ^4}{45}-\frac{304}{9} \zeta_3\right)\nonumber\\
&+T_F^2 n_f^2\left(\frac{4160}{729}+\frac{40 \pi ^2}{81}
  -\frac{224}{27} \zeta_3\right).
\label{eq:gamma-s}
\end{align}
The heavy particle anomalous-dimension coefficients are only
known up to two loops
\begin{equation}
\label{eq:gamma-H}
\begin{aligned}
\gamma_{H,s}^{(0)}&=-2,\\
\gamma_{H,s}^{(1)}&=-C_A\left(\frac{98}{9}-\frac{2\pi^2}{3}+4\zeta_3\right)
+\frac{40}{9} T_F n_f.
\end{aligned}
\end{equation}

\section{Potential corrections}
\label{app:potential}
\subsection{Potential for general spin and colour states}
\label{sec:pot-m2}
In this appendix we provide the projection of the PNRQCD potential
onto spin-singlet and triplet states and obtain the one-loop $1/m^2$
potential for the colour-octet case. The conventions of the
PNRQCD Lagrangian  follow~\cite{Beneke:2013jia}.
The general colour dependence of the quark-antiquark potential can be
decomposed in the two equivalent forms\footnote{Note that $V_S$ is called $V_1$ in~\cite{Beneke:2013jia}.}
\begin{equation}
  \begin{aligned}
    V_{ab;cd} ({\bf p},{\bf p}^\prime) &=
    V_S({\bf p},{\bf p}^\prime) \delta_{ab} \delta_{cd} 
    +V_T ({\bf p},{\bf p}^\prime)T^A_{ab} T^A_{cd}\\
    &=V^{1}({\bf p},{\bf p}^\prime) \frac{1}{N_c}\delta_{cb}\delta_{ad} 
    +V^8({\bf p},{\bf p}^\prime) 2 T^A_{cb}T^A_{ad},
\end{aligned}
\label{eq:project-potential}
\end{equation}
where the potential coefficients in the two conventions are related by
\begin{equation}
  V^R=V_S-D_R V_T,
\label{eq:colour-potential}
\end{equation}
with the colour factor~\eqref{eq:colour-coulomb}.
The potential for arbitrary spin in a given colour channel $R$ is of
the form~\cite{Beneke:2013jia}
\begin{eqnarray}
V^R({\bf p},{\bf p}^{\prime})&=& 
{\cal V}^R_C(\alpha_s)\frac{4\pi D_R\alpha_s}{{\bf
q}^2} -
{\cal V}^R_{1/m}(\alpha_s) \frac{\pi^2 (4\pi)D_R\alpha_s} {m_t|{\bf q}|}  
\nonumber  \\ 
&& -\, {\cal V}^R_{\delta}(\alpha_s)\frac{2\pi D_R \alpha_s}{m_t^2}
\nonumber
 +  {\cal V}_{s}^R(\alpha_s)\frac{\pi D_R
 \alpha_s}{4m_t^2}[\sigma_i,\sigma_j]\otimes[\sigma_i,\sigma_j] 
\nonumber  \\ 
&&   +\,{\cal V}^R_{p}(\alpha_s)\frac{2\pi D_R
 \alpha_s({\bf p}^2+{\bf p}^{\prime 2})}{m_t^2{\bf{q}}^2}
-{\cal V}^R_{hf}(\alpha_s)\frac{\pi D_R \alpha_s}{4m_t^2{\bf q}^2}
 [\sigma_i,\sigma_j]q_j\otimes [\sigma_i,\sigma_k]q_k
\nonumber  \\ 
&&  + {\cal V}^R_{so}(\alpha_s)\frac{3\pi D_R \alpha_s}{2m_t^2 {\bf q}^2}
 \Bigg([\sigma_i,\sigma_j]q_i p_j
 \otimes 1
 -1\otimes [\sigma_i,\sigma_j]q_i p_j\Bigg)
+ \ldots\,.
\label{eq:potentialbeforespin} 
\end{eqnarray}
Here the spin-dependence is written in a tensor-product notation, $a\otimes b$, where 
$a$ ($b$) refers to the spin-matrix on the quark (antiquark). In
addition to~\eqref{eq:potentialbeforespin}, 
also the so-called annihilation contribution from local NRQCD
four-fermion operators must be included for
hadronic top-quark production. This is discussed
in App.~\ref{app:annihilation}.
The potential~\eqref{eq:potentialbeforespin}  can be further decomposed into 
spin-singlet and triplet contributions according to
\begin{equation}
\label{eq:project-spin}
  V^R({\bf p},{\bf p}^{\prime})=\frac{1}{2}V^{R,0}({\bf p},{\bf
    p}^{\prime})\,1\otimes 1
  +\frac{1}{2}V^{R,1}({\bf p},{\bf p}^{\prime}) \,\sigma_i\otimes \sigma_i, 
\end{equation}
where the  spin projection in $d-1$ space dimensions
can be performed as discussed in Section 4.5
of~\cite{Beneke:2013jia}. In this way, one obtains
the colour- and spin-projected potential 
\begin{eqnarray}
\label{eq:potentialafterspin} 
V^{R,S}({\bf p},{\bf p}^{\prime}) &=& 
  \frac{4\pi \alpha_{s}D_R}{{\bf q}^2} \bigg[\,{\cal V}^R_C
  -{\cal V}^R_{1/m}\,\frac{\pi^2\,|\bf{q}|} {m_t}
  +{\cal V}^{R,S}_{1/m^2}\,\frac{{\bf q}^2}{m_t^2}
  +{\cal V}_{p}^R\,\frac{{\bf p}^2+{\bf p}^{\prime \,2}}{2m_t^2}\,
\bigg].
\end{eqnarray}
Additionally, the term $\partial^4/8m_t^3$ in the PNRQCD Lagrangian is
also treated as a potential (cf. Eq.(4.60) of~\cite{Beneke:2013jia}).

We require the Coulomb-potential up to two-loop accuracy, which reads
in four dimensions
\begin{equation}
\label{eq:coulomb}
  {\cal V}_C^R=1+\frac{\alpha_s}{4\pi}\left(a_1+\beta_0
    \ln\frac{\mu^2}{{\bf
        q}^2}\right)+\left(\frac{\alpha_s}{4\pi}\right)^2\left( a_2^R+(2a_1\beta_0+\beta_1)
    \ln\frac{\mu^2}{{\bf q}^2}  +\beta_0^2  \ln^2\frac{\mu^2}{{\bf q}^2}  \right)+\dots
\end{equation}
with the one-loop coefficient
\begin{equation}
  \label{eq:a1}
a_1  
=\frac{31}{9} C_A-\frac{20}{9} n_l T_f.
\end{equation}
The two-loop coefficient depends on the colour representation~\cite{Schroder:1998vy,Kniehl:2004rk},
\begin{equation}
  \label{eq:a2}
\begin{aligned}
    a_2^{1}&=C_A^2\left(\frac{4343}{162}+4\pi^2
      -\frac{\pi^4}{4}+\frac{22}{3}\zeta_3\right)
    -C_A T_F n_f \left(\frac{1798}{81}+\frac{56}{3}\zeta_3\right)\\
    &- C_F T_F n_f\left(\frac{55}{3}-16\zeta_3\right)
    +(T_F n_f)^2\frac{400}{81},\\
    a_2^{8}&=a_2^{1}+C_A^2(\pi^4-12\pi^2).
\end{aligned}  
\end{equation}
The $1/m$ potential for the colour-singlet case is quoted in Section
4.4.2 of~\cite{Beneke:2013jia}. The colour-octet one-loop result can be found
in~\cite{Beneke:2016kvz}, while the corresponding two-loop result is
currently not known. The resulting uncertainty on our predictions is
estimated as discussed in Section~\ref{sec:partonic}.

The spin-dependence in~\eqref{eq:potentialafterspin} enters only
through the $1/m^2$ potential
\begin{align}
  {\cal V}^{R,S}_{1/m^2}=-\frac{1}{2}{\cal V}^R_\delta 
  +\left(v^S(\epsilon)+\frac{1}{2}\right)
  \left[{\cal V}^R_{hf}-(d-1) {\cal V}^R_s \right],
\end{align}
where in $d=4-2\epsilon$ dimensions
\begin{equation}
  v^S(\epsilon)= 
\begin{cases}
  -\frac{1}{2}\epsilon, &  S=0\\
  \frac{2\epsilon^2+\epsilon-4}{6-4\epsilon} ,& S=1
\end{cases}
\end{equation}
At tree level, the potential coefficients in the $1/m^2$ potential are
\begin{align}
  {\cal V}_\delta^{R(0)}&=1, &
  {\cal V}_{hf}^{R(0)}&=1, &
  {\cal V}_{s}^{R(0)}&=0, &
  {\cal V}_{p}^{R(0)}&=1,
\end{align}
so that
\begin{equation}
{\cal V}^{R,S,(0)}_{1/m^2}=v^S(\epsilon).
\end{equation}

The one-loop result for the colour-singlet $1/m^2$ potential is given in Section
4.4.3 of~\cite{Beneke:2013jia}. 
We argue now  that the colour-octet result can be obtained from
the singlet case after an appropriate
adjustment of the colour factors. 
Following~\cite{Beneke:2013jia},   the  one-loop $1/m^2$ potential can be calculated
from an NRQCD computation consisting of a ``hard'' contribution of tree diagrams
with one-loop matching coefficients, and a ``soft'' contribution of one-loop
NRQCD diagrams. 
The soft contributions receive contributions from the diagrams in Fig.~9
of~\cite{Beneke:2013jia}. It can be seen from the NRQCD Feynman rules that the
colour factors are either the same as the box or crossed-box topologies in the
full QCD diagrams, 
\begin{align}
  C^{\text{box}}_{ab;cd} &=(T^AT^B)_{ab}(T^BT^A)_{cd}\,, & 
  C^{\text{c-box}}_{ab;cd} &=(T^AT^B)_{ab}(T^AT^B)_{cd},
\end{align}
or can be written as linear combinations thereof.
 The projection of these colour factors on the
singlet and octet states can be written in the form $C^{\text{box}}\to D_R^2$ and
$C^{\text{c-box}}\to D_R\left(D_R+\frac{C_A}{2}\right)$ so that in both cases
the result for the general representation is obtained from the singlet
result by the replacement $(-C_F)\to D_R$.
There are further contributions from the abelian vertex correction, self-energy
insertions and charge-renormalization counterterms.
These contributions contain explicit factors of $C_F$ but are scaleless, so
they do not contribute to the finite part of the potential.
The hard corrections receive contributions from one-loop matching
coefficients of four-fermion operators, the NRQCD quark-gluon
vertex and the gluon two-point functions.  Since the former arise from
box-diagrams in full QCD, the same replacement of colour factors as in
the soft contribution can be applied, while the colour factors arising
from the coefficients $d_i$ of the vertex and two-point functions  are unmodified.\footnote{Formally this result
  is obtained by adjusting the colour projection of the contribution of
the Wilson coefficients of the four-fermion operators in the expressions for
$\mathcal{V}_{\delta}^{(\text{hard})}$ and  $\mathcal{V}_{s}^{(\text{hard})}$
in~(4.75) and~(4.79) of~\cite{Beneke:2013jia} according
to~\eqref{eq:colour-potential}.}
As result, we obtain the  $1/m^2$ potential for the spin-singlet/triplet and colour-singlet/octet states 
\begin{eqnarray}
{\cal V}_{1/m^2}^{R,S,(1)} &=&
   \Bigg[\bigg(\frac{\mu^2}{\bf{q}^2} \bigg)^{\!\epsilon }-1
   \Bigg]\,
   \frac{1}{\epsilon}\,
   \bigg(\,-\frac{7}{3}D_R-\frac{25+21v^S(\epsilon)}{6}C_A\, + \beta_0\, v^S(\epsilon)\bigg)
\nonumber
\\
&&+\,\Bigg[\left(\frac{{\mu }^2}{m_t^2} \right)^{\!\epsilon}-1
   \Bigg]\, \frac{1}{\epsilon}\,
   \bigg(\,\frac{4C_F+3D_R}{3}+\frac{17+21 v^S(\epsilon)}{6}C_A\,\bigg)\nonumber
\\
&&
  +\bigg(\frac{\mu^2}{\bf{q}^2} \bigg)^{\!\epsilon}\,
        v^{S(1)}_{q}(\epsilon)
  +\bigg(\frac{{\mu }^2}{m_t^2} \bigg)^{\!\epsilon}\,
        v^{S(1)}_{m}(\epsilon),\quad
\label{eq:oneoverm2}
\end{eqnarray}
with
\begin{align}
v^{S,(1)}_{q}(\epsilon) &= \frac{D_R}{3} 
-\frac{7}{4}C_A-\left(\frac{5}{9} C_A
+\frac{20}{9} \,T_F n_f\right) v^S(\epsilon)+O(\epsilon) ,
\nonumber \\
v^{S,(1)}_{m}(\epsilon) &= 
\frac{11\,D_R}{3} + 2C_F-\frac{C_A}{4}
+(6 D_R+4 C_F+3 C_A) v^S(\epsilon)
+\frac{4}{15}\,T_F+\,O(\epsilon),
\end{align}
where the $O(\epsilon)$ contributions do not contribute to the logarithmic
corrections considered here.
 This result agrees with an explicit calculation of the singlet and octet cases~\cite{Wuester:2003}.
 The potential coefficient ${\cal V}_{p}^{R,(1)}$ is the same for both the singlet
 and the octet case.
 
\subsection{Annihilation contribution}
\label{app:annihilation}
The contribution of local four-fermion operators to the NRQCD Lagrangian is
given by
\begin{equation}
  \delta\mathcal{L}_{4f}=
-\delta V_{ab;cd}^{4f}\;
\left(\psi_{c}^{\dagger}\chi_{b} 
\chi^\dagger_{a}
\psi_{d}\right),
\end{equation}
where the labels collectively denote the spin and colour quantum
numbers. This convention agrees with~\cite{Beneke:2016kvz}.  The four-fermion
Lagrangian can be decomposed into contributions with definite spin and colour
quantum numbers, analogously to~\eqref{eq:project-potential}
and~\eqref{eq:project-spin}
\begin{equation}
  \begin{aligned}
- \delta\mathcal{L}_{4f}&=
\frac{\delta V_{4f}^{1,0}}{2N_C}\left(\psi^{\dagger}\chi\right) 
\left(\chi^\dagger\psi\right)
+\delta V_{4f}^{8,0}\;
\left(\psi^{\dagger}T^A\chi\right) 
\left(\chi^\dagger T^A \psi\right) \\
&\quad+\frac{\delta V_{4f}^{1,1}}{2 N_C}
\left(\psi^{\dagger}\sigma^i\chi\right) 
\left(\chi^\dagger \sigma^i \psi\right)
+\delta V_{4f}^{8,1}\;
\left(\psi^{\dagger}T^A\sigma^i \chi\right) 
\left(\chi^\dagger T^A \sigma^i \psi\right).
  \end{aligned}
\end{equation}
The relation to the coefficients $d^c$ introduced
in~\cite{Pineda:1998kj} is given by
\begin{align}
 \delta V_{4f}^{1,0}&=-2N_C \frac{d^c_{ss}}{m_t^2}, &
  \delta V_{4f}^{8,0}&=-\frac{d^c_{vs}}{m_t^2},&
   \delta V_{4f}^{1,1}&=-2N_C \frac{d^c_{sv}}{m_t^2}, &
  \delta V_{4f}^{8,1}&=-\frac{d^c_{vv}}{m_t^2}.
\end{align}

The contributions to the four-fermion operators arising from ``scattering''
diagrams (i.e. diagrams present if $\psi$ and $\chi$ correspond to different
quark flavours) are already included in the PNRQCD $1/m^2$ potential discussed
in appendix~\ref{sec:pot-m2}. For the ``annihilation'' contributions
(i.e. diagrams only present for identical quark flavours) we use the notation
\begin{equation}
    \delta V_{\mathrm{ann}}^{R,S}=\frac{\pi \alpha_s}{m_t^2}\nua.
\end{equation}
The perturbative corrections to the annihilation coefficients up to NLO will be written in the form
\begin{equation}
    \nu_{\mathrm{ann}}^{R,S}(\mu)=
    {\nu_{\mathrm{ann}}^{R,S}}^{(0)}+\frac{\alpha_s}{4\pi}   
    {\nu_{\mathrm{ann}}^{R,S}}^{(1)}(\mu)+\mathcal{O}(\alpha_s^2).
\end{equation}
The scale-dependence of the NLO coefficient only enters through the running of
$\alpha_s$:\footnote{Note that the results
  in~\cite{Pineda:1998kj} are given in terms of $\alpha_s^{n_l+1}(\mu)$
  whereas we use $n_l$ active quark flavours in the running.}
\begin{equation}
    {\nu_{\mathrm{ann}}^{R,S}}^{(1)}(\mu)=
   {\nu_{\mathrm{ann}}^{R,S}}^{(1)}(2m_t)-2\beta_0\ln\left(\frac{2m_t}{\mu}\right)
    {\nu_{\mathrm{ann}}^{R,S}}^{(0)}.
\end{equation}
At LO, the
only non-vanishing correction is
\begin{equation}
    {\nu_{\mathrm{ann}}^{8,1}}^{(0)}=1,
\end{equation}
which is relevant for the quark-antiquark channel.
At NLO there are also non-vanishing annihilation contributions in the
spin-singlet channel, which are relevant for gluon-induced top production~\cite{Pineda:1998kj}:
\begin{equation}
\begin{aligned}
    {\nu_{\mathrm{ann}}^{8,1}}^{(1)}(2m_t)&=-\frac{20}{9}T_F n_l-\frac{32}{9} T_F
    -16 C_F +C_A\left(\frac{109}{9}+\frac{22}{3}\ln 2\right),\\
   {\nu_{\mathrm{ann}}^{1,0}}^{(1)}(2m_t)&= - 8C_F(1-\ln 2),\\
     {\nu_{\mathrm{ann}}^{8,0}}^{(1)}(2m_t)&= -4C_F\left(-\frac{3C_A}{2}+4C_F\right)(1-\ln 2).
\end{aligned}
\end{equation}
Here we have neglected imaginary parts, which contribute to toponium decay
into light hadrons and should not be taken into account for the total cross
section for $pp\to t\bar t\to b\bar b W^+W^-$. 

The corrections to the NNLO Green function due to an insertion of an
annihilation correction is~\cite{Beneke:2016kvz}
\begin{equation}
  \delta_{\mathrm{ann}} G_{R}^{(1)S}   
   = - \frac{\pi \alpha_s}{m_t^2}{\nua}^{(0)}
   \left[ G_{R}^{(0)}(0,0;E) \right]^2,  
   \label{eq:dG-ann2}
\end{equation}
with the resulting correction to the potential function at
$\mathcal{O}(\alpha_s^2)$ 
\begin{equation}
  \Delta_{\mathrm{ann}} J_{R}^{S(2)}(E)= J^{(0)}(E)\alpha_s^2(\mu)
  \frac{D_R}{4}{\nua}^{(0)} (2L_E+1).
\end{equation}
The annihilation
corrections to the N$^3$LO Green function are obtained from the master formula
 in~\cite{Beneke:2013jia} as
\begin{equation}
  \delta_{\mathrm{ann}} G_{R}^{(2)S}   
   = - \frac{\alpha_s^2}{4m_t^2}{\nua}^{(1)}(\mu)
   \left[ G_{R}^{(0)}(0,0;E) \right]^2 
   - \frac{\pi \alpha_s}{m_t^2}{\nua}^{(0)}
   2 G_{R}^{(0)}(0,0;E) G_{R}^{(1)}(0,0;E),
   \label{eq:dG-ann3}
\end{equation}
where $G_{R}^{(1)}(0,0;E)$ is the Green function with the insertion of one NLO
Coulomb potential. The second term has the same form as the double insertion
of the NLO Coulomb potential and a delta potential treated
in~\cite{Schuller:2008rxa} and included in the calculation
of~\cite{Beneke:2013jia,Beneke:2015kwa,Beneke:2016kkb}, so the annihilation
correction can be obtained from these results. The logarithmic
$\mathcal{O}(\alpha_s^3)$ corrections to the potential function are given by
\begin{align}
   \Delta J_{R,\text{ann}}^{S(3)}(E)&= J^{(0)}(E)\frac{\alpha_s^3(\mu)}{4\pi}
  \frac{D_R}{2} \Biggl[\beta_0{\nua}^{(0)}
  \left(L_E^2- (2L_E+1)\ln\left(\frac{2m_t}{\mu}\right)\right)\nonumber\\
    &\qquad +({\nua}^{(1)}(2m_t)+a_1 {\nua}^{(0)})L_E
   \Biggr].
\end{align}

\section{Scaling functions}
\label{app:scaling}
In this appendix we collect the numerical expressions for the threshold expansion
of the scaling functions $ f^{(3,i)}_{pp'(R)}$ with $i\geq 1$ parameterizing
the factorization-scale dependence, and the constants $C^{(3)}_{q\bar q(8)}$
in the scaling functions $f^{(3,0)}_{pp'(R)}$.  The results of the scaling
functions with $i=2,3$ are already fully determined at NNLL accuracy. We
include the expressions given already in Appendix C of~\cite{Beneke:2011mq}
for completeness. 
The constant terms in the scaling functions depend on the parameters $k_s$,
$k_h$, and $k_C$ in the unphysical soft, 
hard, and Coulomb scales $\mu_s=k_s m_t v^2$,
$\mu_h=k_h m_t$, $mu_C=k_C m_t v$.

\subsection{Quark-antiquark channel}
The constant in the threshold expansion in the
quark-antiquark channel is given by
\begin{align}
  C^{(3)}_{q\bar q(8)}(k_h,k_s)=&
  6951.36+202.2716 \ln ^6k_h-1574.459 \ln ^5k_h+3123.481
   \ln ^4k_h\nonumber\\
    &+3828.93 \ln ^3k_h-6813.31 \ln^2k_h-39816.5 \ln k_h-202.2716 \ln^6k_s
    \nonumber\\
    &-1167.785 \ln ^5k_s-1220.5 \ln^4k_s+6305.0 \ln ^3k_s+15434.0 \ln^2k_s 
    \nonumber\\
    &
    +6050.0 \ln k_s.
\end{align}
The scaling functions of the scale-dependent contributions are
\begin{align}
  f^{(3,3)}_{q\bar q(8)}=&
-12945.4 \ln^3\beta+39223.6 \ln^2\beta-28867.1 \ln\beta+1140.45,\\[0.2cm]
f^{(3,2)}_{q\bar q(8)}=&\frac{1}{\beta }
\left(-2994.51 \ln^2\beta+5287.18 \ln\beta-1005.67\right)+38836.1 \ln^4\beta
-111164\ln^3\beta \nonumber\\
    &+78770 \ln ^2\beta-3383.32 \ln\beta-3697.7,\\[0.2cm]
f^{(3,1)}_{q\bar q(8)}=&
\frac{1}{\beta^2}\left(-153.93\ln\beta+56.8546\right)
+\frac{1}{\beta}
(5989.02 \ln ^3\beta-7733.19 \ln ^2\beta-2669.2 \ln \beta\nonumber\\
&+2272.92) -38836.1 \ln^5\beta+109310. \ln^4\beta
-78403.7 \ln ^3\beta-37732.\ln^2\beta\nonumber\\
&+36036.4 \ln\beta+38614.
\end{align}
The constant terms in these functions are given for completeness but are not
included in our default approximation.
These results agree with those obtained at NNLL, with the exception of the
coefficient of the $\log\beta$-term and the constant in $f^{(3,1)}_{q\bar q(8)}$, which are a
new result.

\subsection{Gluon fusion colour-singlet channel}
The constant in the scaling function for the colour-singlet gluon-gluon
initial state is given by
\begin{align}
  C^{(3)}_{gg(1)}(k_h,k_s)=&-23367.7+2304.0 \ln ^6k_h-12526.07 \ln ^5k_h+12833.21
   \ln ^4k_h\nonumber\\
    &-3051.61 \ln ^3k_h+24684.5  \ln^2k_h+ 117774.0\ln k_h-2304.0 \ln^6k_s\nonumber\\
    &-1845.80 \ln ^5k_s+25496.0 \ln^4k_s+58820.0\ln ^3k_s
    +6290.0\ln^2k_s\nonumber\\
    &-49700.0 \ln k_s-21543.91 \ln k_C.
\end{align}
The Coulomb-scale dependence arises here because the P-wave contributions are
only included with the LO potential function.  The NLL treatment of the P-wave
corrections leads also to some spurious contributions due to incomplete
cancellations between the P-wave contribution and contributions arising from
the  term  involving $h^{(0),R}_{pp',P}$ in the
two-loop constant~\eqref{eq:cbar}, which are entering in the S-wave
contribution that is treated at N$^3$LL. Since we only use the constant as an
uncertainty estimate, we do not attempt to fix this issue.

The scaling functions of the scale-dependent contributions read
\begin{align}
f^{(3,3)}_{gg(1)}=&\,
- 147456\ln^3\beta
+206398\ln^2\beta+10843.3 \ln\beta-50289.8 
\,,\\[0.2cm]
f^{(3,2)}_{gg(1)}=&\,\frac{1}{\beta }
\left(121278\ln^2\beta-58112.2 \ln\beta-43685.6\right)
+442368\ln^4\beta-448309\ln^3\beta\nonumber\\
&
-297750\ln^2\beta+322713\ln\beta-52925.9 
,\label{eq:f32gg1}\\[0.2cm]
f^{(3,1)}_{gg(1)}=&\,\frac{1}{\beta^2}\left(
- 22166\ln \beta-8283.49\right)
+\frac{1}{\beta}(-242555\ln^3\beta-51902.1 \ln^2\beta
+217024\ln \beta\nonumber\\
&
-29709.3)-442368\ln^5\beta+300977 \ln^4\beta+563908\ln
^3\beta-618319\ln^2\beta\nonumber\\
&
+132635\ln\beta-16096.3\,.
 \end{align}
\subsection{Gluon fusion colour-octet channel}
The constant in the scaling function for the colour-octet gluon-gluon
initial state is
\begin{align}
C^{(3)}_{gg(8)}(k_h,k_s)=&-420.781+2304.0 \ln ^6k_h-9070.07 \ln ^5k_h
-288.71 \ln^4k_h\nonumber\\
&+31337.14 \ln ^3k_h-10948.9 \ln^2k_h-33050.8  \ln k_h-2304.0 \ln^6k_s\nonumber\\
&-5301.8 \ln ^5k_s+18005.0 \ln^4k_s+74880.0 \ln ^3k_s+60510.0\ln^2k_s\nonumber\\
&-22200.0\ln k_s-336.6236 \ln k_C.
\end{align}
The scaling functions for the factorization-scale dependent terms are
\begin{align}
f^{(3,3)}_{gg(8)}=&f^{(3,3)}_{gg(1)},
\\[0.2cm]
f^{(3,2)}_{gg(8)}=&\,\frac{1}{\beta }
\left(-15159.7 \ln ^2\beta +7264.03 \ln\beta+5460.7\right)
+ 442368 \ln ^4\beta-558901 \ln ^3\beta \nonumber\\
&
-100578\ln^2\beta+284537 \ln\beta-88163.8 , \\[0.2cm]
f^{(3,1)}_{gg(8)}=&\,
\frac{1}{\beta ^2}\left(-346.343 \ln\beta-129.429\right)
+\frac{1}{\beta }
(30319.4 \ln^3\beta-1092.09 \ln^2\beta-21802.5 \ln\beta\nonumber\\
&
+5903.022)
-442368\ln^5\beta+522161\ln ^4\beta+227359\ln^3\beta
-748536.1\ln^2\beta\nonumber\\
&+351491\ln\beta+1103.01 .
\label{eq:f31gg8}
\end{align}

\bibliography{bib_resum}

\end{document}